\documentclass[sigplan,10pt,screen]{acmart}
\renewcommand\footnotetextcopyrightpermission[1]{} 
\def\BibTeX{{\rm B\kern-.05em{\sc i\kern-.025em b}\kern-.08emT\kern-.1667em\lower.7ex\hbox{E}\kern-.125emX}}

\usepackage{epstopdf}

\usepackage{epsfig,endnotes}
\newcommand{\secref}[1]{\S\ref{#1}} 

\usepackage{xcolor} 
\usepackage{algorithm} 
\usepackage{algorithmic}

\usepackage{subcaption} 
\usepackage[labelformat=simple]{subcaption}

\usepackage{amssymb}
\usepackage{pifont}
\newcommand{\cmark}{\ding{51}}%
\newcommand{\xmark}{\ding{55}}%

\usepackage{array} 
\newcolumntype{P}[1]{>{\centering\arraybackslash}p{#1}}
\newcolumntype{M}[1]{>{\centering\arraybackslash}m{#1}}

\usepackage{tikz}
\usepackage{xcolor}
\newcommand{\ballnumber}[1]{\tikz[baseline=(myanchor.base)] \node[circle,fill=.,inner sep=1pt] (myanchor) {\color{-.}\bfseries\footnotesize #1};}

\usepackage{enumitem} 
\setlist{nolistsep} 

\usepackage{multirow}

\begin{document}


\hyphenpenalty = 5000
\tolerance = 1000

\title{Efficient Similarity-aware Compression to Reduce Bit-writes in Non-Volatile Main Memory for Image-based Applications}

\author{Zhangyu Chen, Yu Hua, Pengfei Zuo, Yuanyuan Sun, Yuncheng Guo}
\affiliation{%
	\institution{Huazhong University of Science and Technology}}

\begin{abstract}
Image bitmaps have been widely used in in-memory applications,
which consume lots of storage space and energy. Compared
with legacy DRAM, non-volatile memories (NVMs) are suitable for
bitmap storage due to the salient features in capacity and
power savings. However, NVMs suffer from higher
latency and energy consumption in writes compared with reads. Although compressing data
in write accesses to NVMs on-the-fly reduces the bit-writes in NVMs, 
existing precise or approximate compression
schemes show limited performance improvements for data of bitmaps, 
due to the irregular data patterns and variance in data. We observe that the data
containing bitmaps show the pixel-level similarity
due to the analogous contents in adjacent pixels. By exploiting the pixel-level
similarity, we propose SimCom, an efficient similarity-aware compression
scheme in hardware layer, to compress data for
each write access on-the-fly. The idea behind SimCom is to compress
continuous similar words into the pairs of base words with runs. 
With the aid of domain knowledge of
images, SimCom adaptively selects an appropriate compression mode
to achieve an efficient trade-off between image quality and memory performance.
We implement SimCom on GEM5 with NVMain and evaluate the performance
with real-world workloads. Our results demonstrate that SimCom reduces 33.0\%, 34.8\%
write latency and saves 28.3\%, 29.0\% energy than state-of-the-art FPC and BDI with minor
quality loss of 3\%.
\end{abstract}

%
%


%

\settopmatter{printfolios=true, printacmref=false} 
%
\maketitle

\section{Introduction}



Images have been widely used and stored in large-scale storage systems.
Compared with encoded bits, image bitmaps (also called raster images) 
contain pixel-level information
(e.g., image features~\cite{DBLP:journals/ijcv/Lowe04,
DBLP:conf/iccv/RubleeRKB11}) used by various applications (e.g., image
processing, computer vision, and machine learning). 
In order to preserve the pixel-level information for these applications, 
images need to be stored as bitmaps in the main 
memory~\cite{DBLP:conf/iccad/ZhaoXCZ17}. However, the storage 
of bitmaps demands a large amount of memory and energy in DRAM.
Unlike it, Non-Volatile Memories (NVMs) are more suitable for bitmap 
storage due to the high density, DRAM-scale read latency, byte-addressability, 
and near-zero standby power~\cite{DBLP:conf/usenix/DavidDGZ18}. 
Using NVM-based main memory for
image-based applications is promising to improve the efficiency 
in terms of storage density and energy consumption, thus improving 
the overall system performance.

NVMs offer high-density memory for image bitmaps, and the writes of
bitmaps often cause performance degradation in NVM systems.
NVMs, such as Phase Change
Memory (PCM) and Resistive RAM (ReRAM), suffer from high write latency and
non-negligible write energy consumption~\cite{DBLP:conf/isca/ZhouZYZ09,
DBLP:conf/isca/LeeIMB09,DBLP:conf/hpca/LiZL13a,DBLP:conf/hpca/YueZ13,
DBLP:conf/osdi/Zuo0W18,DBLP:conf/sosp/XuZMGBSSR17,
DBLP:conf/sosp/ConditNFILBC09}.

\begin{table}[!ht]
  \vspace{-0.1cm}
  \centering
  \caption{The ratios of written data containing image 
  bitmaps. \emph{(The granularity of data in a write
  access is equal to the cache block size, e.g., 64 bytes.)}}
  \vspace{-0.2cm}
  \label{tab:BitmapBlockRatio}
  \begin{tabular}{|c|c||c|c|}
    \hline
    \textbf{Workloads} & \textbf{Ratio} & \textbf{Workloads} & \textbf{Ratio} \\ \hline
    jpeg & 97.6\% & 2dconv & 99.5\% \\ \hline
    sobel & 99.3\% & debayer & 25.3\% \\ \hline
    kmeans & 27.2\% & histeq & 99.1\% \\
    \hline
  \end{tabular}
  \vspace{-0.2cm}
\end{table}

\begin{figure}[!ht]
  \vspace{-0.2cm}
  \centering
  \includegraphics[width=0.4\textwidth]{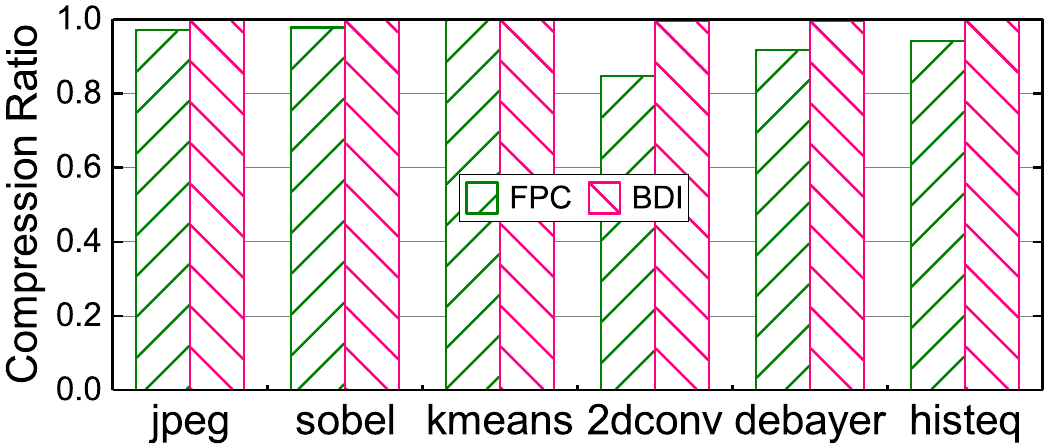}
  \vspace{-0.2cm}
  \caption{\label{fig:BitmapCompressionRatio}The compression ratios of data writes
  containing image bitmaps using FPC and BDI.}
  \vspace{-0.1cm}
\end{figure}

In order to address the write-inefficiency problem, 
an intuitive solution is to compress
images via software-layer coding algorithms (e.g., JPEG 
XR~\cite{dufaux2009jpeg} and JPEG~\cite{DBLP:journals/cacm/Wallace91})
before writing data into NVMs. 
Although software-layer image compression schemes can decrease 
the image sizes, they cause high overheads due to the high complexity. 
Many image-based applications need to access and manipulate 
bitmaps in pixels, thereby being unable to 
operate on compressed images. For example, the kernel in the sobel 
algorithm~\cite{DBLP:journals/dt/YazdanbakhshMEL17} is used to 
read and update the pixels one by one.
Software-layer compression/decompression before each access to images
causes additional latency and decreases the system performance.
Moreover, the intermediate bitmaps (decompressed for processing in
image-based applications) still cause lots of writes to NVMs.
For these applications, software-layer compression can generate
more writes (i.e., compressed images and intermediate bitmaps)
compared with simply storing bitmaps in NVMs.
In the meantime, it's impractical to implement image compression 
algorithms in the hardware layer (e.g., memory controllers) 
due to the high complexity and the small
data size (e.g., 64 B) in each write access.

Recent works propose data compression inside the NVM module controller
to reduce the 
bit-writes for each write access 
on-the-fly~\cite{DBLP:conf/micro/HongNABKH18,DBLP:conf/hpca/PalangappaM16,
DBLP:conf/IEEEpact/PekhimenkoSMGKM12,DBLP:conf/nanoarch/DgienPHLM14}.
These hardware-layer compression schemes partition the 
data into words. Partitioned words are 
compressed in the NVM module controller using general-purpose data 
patterns, e.g., frequent patterns in Frequent Pattern Compression 
(FPC)~\cite{DBLP:conf/nanoarch/DgienPHLM14} and
base words with small deltas in Base Delta Immediate 
(BDI)~\cite{DBLP:conf/IEEEpact/PekhimenkoSMGKM12}.
After compression, compressed data are written to NVMs, 
thus efficiently reducing the bit-writes in NVMs.
However, for write accesses of bitmaps, the partitioned words
are hard to match the frequent patterns designed for general 
applications or satisfy the narrow value constraint of BDI due to 
the large variance, which results in high compression ratios 
(compressed data size relative to uncompressed data size).
We evaluate the compression ratios of FPC and BDI using six 
image-based workloads (\secref{sec:ExperimentalSetting}). 
The numbers in Table~\ref{tab:BitmapBlockRatio} denote the percentage 
of write accesses to NVMs containing bitmaps in each workload.
The writes of bitmaps account for a large portion of NVM writes.
However, as shown in Figure~\ref{fig:BitmapCompressionRatio},
the average compression ratios of FPC and BDI are 94.2\%
and 99.8\%, which means most data writes of image bitmaps obtain
poor compression performance and even become incompressible using 
precise compression schemes.

Recent research explored the approximate storage for images, since
images tolerate minor inaccuracies~\cite{DBLP:conf/asplos/GuoSCM16,
DBLP:conf/iccad/ZhaoXCZ17,DBLP:conf/micro/MiguelAJJ16}.
The approximate image storage proposed by Guo
et al.~\cite{DBLP:conf/asplos/GuoSCM16} leverages the
entropy and error correction requirement differences in the 
encoded bits of compressed images. However, the differences 
don't exist in raw data (i.e., bitmaps). Recent
works~\cite{DBLP:conf/iccad/ZhaoXCZ17,DBLP:conf/micro/MiguelAJJ16} 
exploit the inter-block similarity to provide approximate storage 
for bitmaps. However, 
searching for similar data in NVMs during each write 
access incurs extra latency and hardware overheads.
Since a large portion of data to be written are approximable
(as shown in Table~\ref{tab:BitmapBlockRatio}), it is possible to improve the
memory performance by approximately compressing the data 
on-the-fly before writing to NVMs. In order to
efficiently reduce the bit-writes of bitmaps in NVM systems, 
there are two challenges for data compression.


\textbf{Irregular Data Patterns}.
Data writes containing bitmaps are hard to match 
data patterns in existing compression schemes.
Bitmaps consist of the bits of each pixel, and a typical pixel 
consists of 3 bytes (the pixel size can be 
different \secref{sec:BitmapBackground}).
Since the pixel size in common bitmaps (e.g., 3 B) is not the same
as the word size in conventional compression schemes (e.g., 4 B), there
is significant variance in partitioned words. Besides, the value of
each word depends on the contents of bitmaps.
Therefore, the partitioned words in conventional schemes show irregular
data patterns, which lead to poor compression performance.


\textbf{Bitmap Format Variance}.
When multiple applications (or threads) are running on top
of NVM systems with different bitmap formats, write accesses
to NVMs contain different data layouts. In the meantime, 
the persistence order is determined by the cache replacement 
policy, which is different from the program 
order~\cite{DBLP:conf/osdi/Zuo0W18,DBLP:conf/micro/ShinTTS17}.
Due to the reordering, it's challenging to determine 
the bitmap format for each write access.
Data compression designed for one bitmap format
may fail in others due to the significant changes in data patterns.



Bidirectional precision
scaling~\cite{DBLP:conf/islped/RanjanRRR17} partitions data
using annotated word size and conducts approximate compression 
for approximable data. Specifically, it approximately truncates
Most Significant Bits (MSB) and Least Significant Bits (LSB) 
of error-tolerant data within the accuracy constraint.
However, the pixel value in bitmaps is often stored using the
smallest data type, in which identical MSBs are usually unavailable in bitmaps. 
Moreover, indiscriminately truncating LSBs reduces the color depth and
causes noticeable quality degradation.

To address the above two challenges,
we propose SimCom, an efficient
similarity-aware compression scheme inside the NVM module controller, 
to reduce the bit-writes of bitmaps into NVMs, 
thus improving the write performance and decreasing the
energy consumption of NVMs for image-based applications.
For the first challenge, we leverage the pixel-level similarity in data writes
of bitmaps and only write a \textbf{base word} (the representative word for a 
group of continuous similar words) with a \textbf{run} (the number of 
words in the group) for each group of continuous similar words, 
which eliminates the writes of similar words in NVMs.
For the second challenge, SimCom executes compression modes in parallel
and adaptively selects an efficient compression mode 
without programmer annotations on image metadata. 

While we use RGB color model to illustrate the compression for
data writes of bitmaps, SimCom can be adapted to other 
color models that show the pixel-level similarity (e.g., YUV and YCbCr).
In addition to images, SimCom also works for other error-tolerant data, 
as long as these data consist of data units of fixed size and
similarity exists in adjacent units, e.g., videos and audio signals.

In SimCom, we make the following contributions:

\begin{itemize}
  \item \textbf{Similarity-aware Compression.}
  By leverage the pixel-level similarity, we develop an efficient 
  approximate data compression scheme in
  hardware layer to reduce the bit-writes of image bitmaps in NVMs on-the-fly.

  \item \textbf{Adaptiveness.}
  With the domain knowledge of bitmaps, we propose an adaptive scheme
  to perform approximate compression without prior knowledge about data formats.
  SimCom eliminates the annotations on the data types and widths of bitmaps.

  \item \textbf{System Implementation.} 
  We have implemented the prototype of SimCom in GEM5 with NVMain and conducted experiments with
  read-world workloads in various domains. Results show that compared with state-of-the-art
  works, SimCom achieves average 33.0\%, 34.8\% write latency reduction
  and 28.3\%, 29.0\% energy savings over FPC and BDI with 3\% quality loss.
\end{itemize}


\section{Background and Motivation}
\label{sec:Background}

\subsection{Image Bitmap}
\label{sec:BitmapBackground}

\indent
\textbf{Structure Organization}.
An image bitmap is a pixel storage structure containing the bits for
each pixel color. The bits of a pixel color 
consist of primary colors. A channel in a bitmap is an image-size 
array of one primary color in each pixel. A typical bitmap consists 
of 3 channels (e.g., red, green, and blue). For each pixel, the number 
of bits per channel is 8. Some bitmaps contain an optional 
channel, \emph{alpha channel}, to store transparency 
information~\cite{DBLP:conf/siggraph/PorterD84,
DBLP:journals/tog/Duff17}. We use channel count 
(\textbf{CC}) to represent the number of channels in a bitmap and bits 
per channel (\textbf{BPC}) to denote the number of bits per channel.

\textbf{Quality Metric}.
Root-mean-square error (RMSE) is an objective metric to measure 
the quality of an image, which accounts for the difference of each 
pixel compared with a baseline image. 
The RMSE of image $\hat{y}$ with respect to baseline image
$y$ is calculated using Equation~\ref{eq:RMSE}. $m$ denotes the number of
pixels in each image. The value of RMSE ranges from 0 to 1 and the lower
value is better performance. We use RMSE to measure the output quality of relaxed images 
like prior works~\cite{DBLP:conf/islped/RanjanRRR17,
DBLP:journals/dt/YazdanbakhshMEL17}.

\vspace{-0.3cm}
\begin{equation}
    RMSE=\sqrt{\frac{1}{m}\sum{_{i=0}^{m-1}(y_i - \hat{y}_i)^2}} \label{eq:RMSE}
\end{equation}
\vspace{-0.4cm}

\subsection{Bit-write Reduction in NVMs}


To address the high latency and energy consumption in write operations, 
bit-write reduction techniques are widely 
used in NVM-based main memory~\cite{DBLP:conf/iscas/YangLKCLY07,
DBLP:conf/micro/ChoL09,DBLP:conf/IEEEpact/PekhimenkoSMGKM12,
DBLP:conf/hpca/YueZ13,DBLP:conf/nanoarch/DgienPHLM14,
DBLP:conf/hpca/PalangappaM16,DBLP:conf/date/XuFHTLL18,
DBLP:conf/date/GuoHZ18,DBLP:conf/dac/PalangappaM18}. 
Related schemes include 
data encoding~\cite{DBLP:conf/iscas/YangLKCLY07,DBLP:conf/micro/ChoL09,
DBLP:conf/hpca/JacobvitzCS13}, 
data compression~\cite{DBLP:conf/IEEEpact/PekhimenkoSMGKM12,
DBLP:conf/nanoarch/DgienPHLM14,DBLP:conf/date/GuoHZ18}, and their 
combinations~\cite{DBLP:conf/hpca/PalangappaM16,DBLP:conf/date/XuFHTLL18,
DBLP:conf/dac/PalangappaM18}. Before writing data into NVMs, compression 
schemes decrease the size of data to be written by data compression. 
Compressed data are decompressed for read accesses.
Data encoding schemes are used to reduce the bit flips in write 
operations. Encoding technologies can be leveraged to encode the 
compressed data for energy efficiency~\cite{DBLP:conf/hpca/PalangappaM16}
and lifetime improvement~\cite{DBLP:conf/date/XuFHTLL18}.

\begin{figure}[!ht]
    \vspace{-0.4cm}
    \centering
    \includegraphics[width=0.46\textwidth]{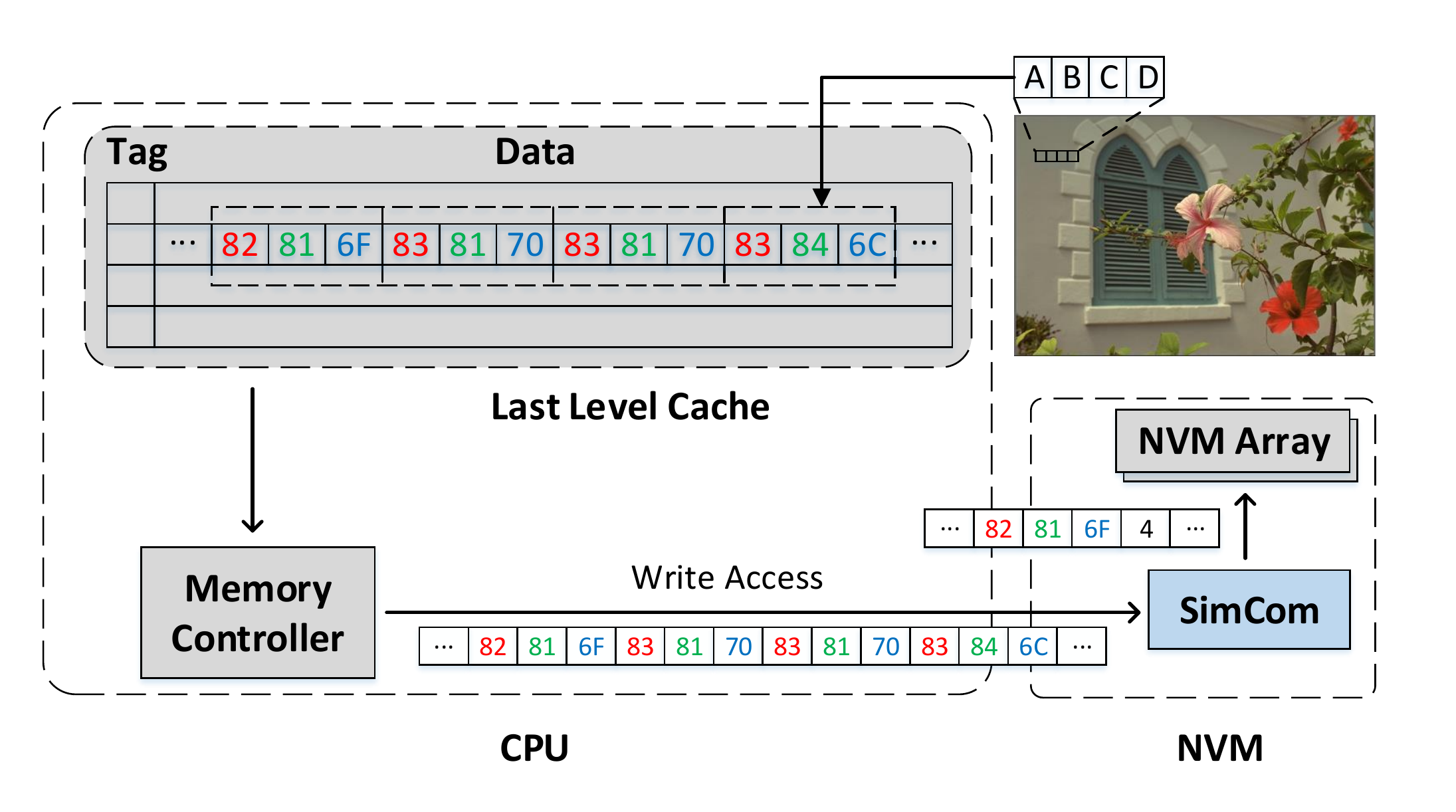}
    \vspace{-0.2cm}
    \caption{\label{fig:PixelSimilarity}An example of leveraging 
    pixel-level similarity to compress data writes.}
    \vspace{-0.4cm}
\end{figure}

\subsection{Approximate Storage}

Approximate storage leverages the error-tolerance of approximable 
data to slightly relax the accuracy constraints for improvement
in performance, data density, lifetime, and energy efficiency. 
Approximable data are interpreted as the data tolerating minor
inaccuracies, which are represented as image bitmaps in the context
of this paper. For error-tolerant applications, typical 
approximation consists of three steps: identification of approximable data, 
approximate techniques, and quality control. Before execution, error-tolerant data 
should be separated from raw application data, which is accomplished by programmer 
annotations~\cite{DBLP:conf/asplos/LiuPMZ11,DBLP:conf/pldi/SampsonDFGCG11,
sampson2015accept,DBLP:conf/micro/MiguelAMJ15,DBLP:conf/micro/MiguelAJJ16,
DBLP:conf/date/RanjanVPV0R17} and domain knowledge~\cite{DBLP:conf/asplos/GuoSCM16,
DBLP:conf/asplos/JevdjicSCM17}. For error-tolerant data, traditional 
guarantees for accuracy in the storage systems are relaxed 
for gains in memory performance and efficiency. Existing approximate techniques
include decreasing refresh 
rate~\cite{DBLP:conf/asplos/LiuPMZ11} and lowering 
voltage~\cite{DBLP:conf/asplos/EsmaeilzadehSCB12} in DRAM, using worn blocks and skipping
program-and-verify iterations in Multi-Level Cell (MLC) 
PCM~\cite{DBLP:conf/micro/SampsonNSC13}, associating similar cache blocks with the same
tag entry~\cite{DBLP:conf/micro/MiguelAMJ15,DBLP:conf/micro/MiguelAJJ16}, 
and utilizing selective error correction code~\cite{DBLP:conf/asplos/GuoSCM16,
DBLP:conf/asplos/JevdjicSCM17}. Given accuracy constraints, in order to obtain the results, 
we need to select appropriate approximation 
parameters~\cite{DBLP:conf/micro/MiguelAMJ15,DBLP:conf/iccad/ZhaoXCZ17} to achieve suitable 
trade-off between output quality and performance. The 
parameters can be inferred dynamically by monitoring the intermediate 
results~\cite{DBLP:conf/pldi/BaekC10,DBLP:conf/asplos/SamadiJLM14}, using the input
features~\cite{DBLP:conf/asplos/SuiLFP16}, and tuning with canary 
inputs~\cite{DBLP:conf/pldi/LaurenzanoHSMMT16,DBLP:conf/usenix/XuKKBMMB18}.

\subsection{Pixel-level Similarity}
\label{sec:IntraBlockSimilarity}


Pixel-level similarity is interpreted as the similarity of words in 
the data of a write access to NVMs. Instead of fixed 4-byte word size,
the data in SimCom are partitioned at pixel-level granularity (e.g., 3 bytes
for RGB format, more details are available in \secref{sec:UniformPartition}). 
In a bitmap, 
each pixel describes the color of a tiny point of the image. 
Hence, adjacent pixels tend to have similar contents. 
As shown in Figure~\ref{fig:PixelSimilarity}, the contents of 
adjacent pixels A, B, C, and D are similar. For the storage of an image 
bitmap, the contents usually are mapped to a continuous region in memory
and have continuous addresses in the address space. When a write access
of bitmap is issued to NVM module and we partition the data at the 
boundaries of pixels, partitioned words are possible 
to be similar due to the analogous contents in adjacent pixels.
This paper proposes to leverage
the pixel-level similarity in data for approximate compression,
thus reducing the data size and improving the memory performance.

\begin{figure}[!ht]
\vspace{-0.2cm}
\centering
\includegraphics[width=0.4\textwidth]{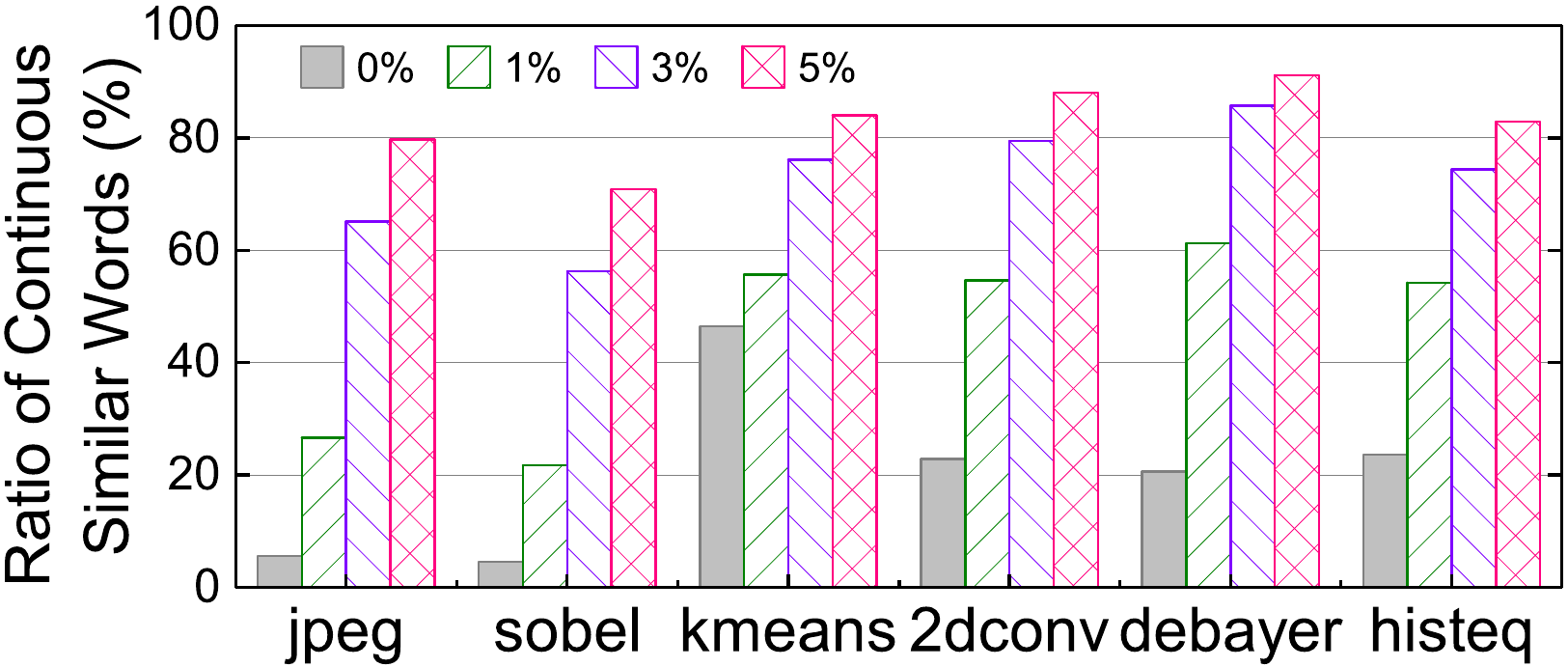}
\vspace{-0.2cm}
\caption{\label{fig:IntraSimilarity}The ratios of continuous similar words in
approximable data with different error thresholds.}
\vspace{-0.2cm}
\end{figure}

We have conducted experiments to verify the pixel-level similarity 
in write accesses to NVMs
by recording continuous similar words in \emph{approximable data}, i.e.,
data containing image bitmaps. 
The similarity metric is described in \secref{sec:SimilarityMetric}.
Approximable data is partitioned at the pixel boundaries (it is possible to
not partition at the boundaries, more details in \secref{sec:UniformPartition}).
In order to present a conservative estimation of pixel-level similarity
in data writes, we filter out the bytes that do not form complete pixels.
Error thresholds range from 0\% (precise) to 100\% 
(maximal approximation), which denotes approximation degrees.
Continuous similar words are interpreted as a group of sequential words, 
in which any two words are similar. The details of experimental 
settings are described in \secref{sec:ExperimentalSetting}.
Figure~\ref{fig:IntraSimilarity} shows the percentage of continuous
similar words in approximable data with different error thresholds. 
When we increase the error threshold, the ratio of 
continuous similar words increases up to 82.8\% on average.

The pixel-level similarity is common in bitmaps due to two reasons: 
(1) The changes in the contents of images
are generally slight. For example, most backgrounds of images consist of similar colors
and lack of abrupt changes. (2) The resolution of images is high. 
With higher resolution for advanced sensors and application requirements, 
the number of pixels corresponding to one item increases and the
difference between two adjacent pixels decreases. The common similarity 
of pixels in images offers the opportunity for approximate compression.

It is worth noting that even when the error threshold is 0\%, the ratio 
of continuous similar words is still more than 4.5\% and up to 46.5\%. 
The substantial similarity in images motivates us to exploit the 
pixel-level similarity for more bit-write reduction in NVMs.

\section{Similarity-aware Data Compression}
\label{sec:ApproximateCompression}

\subsection{Design Overview}
\label{sec:DesignOverview}

\begin{figure}[!ht]
    \centering
    \includegraphics[width=0.44\textwidth]{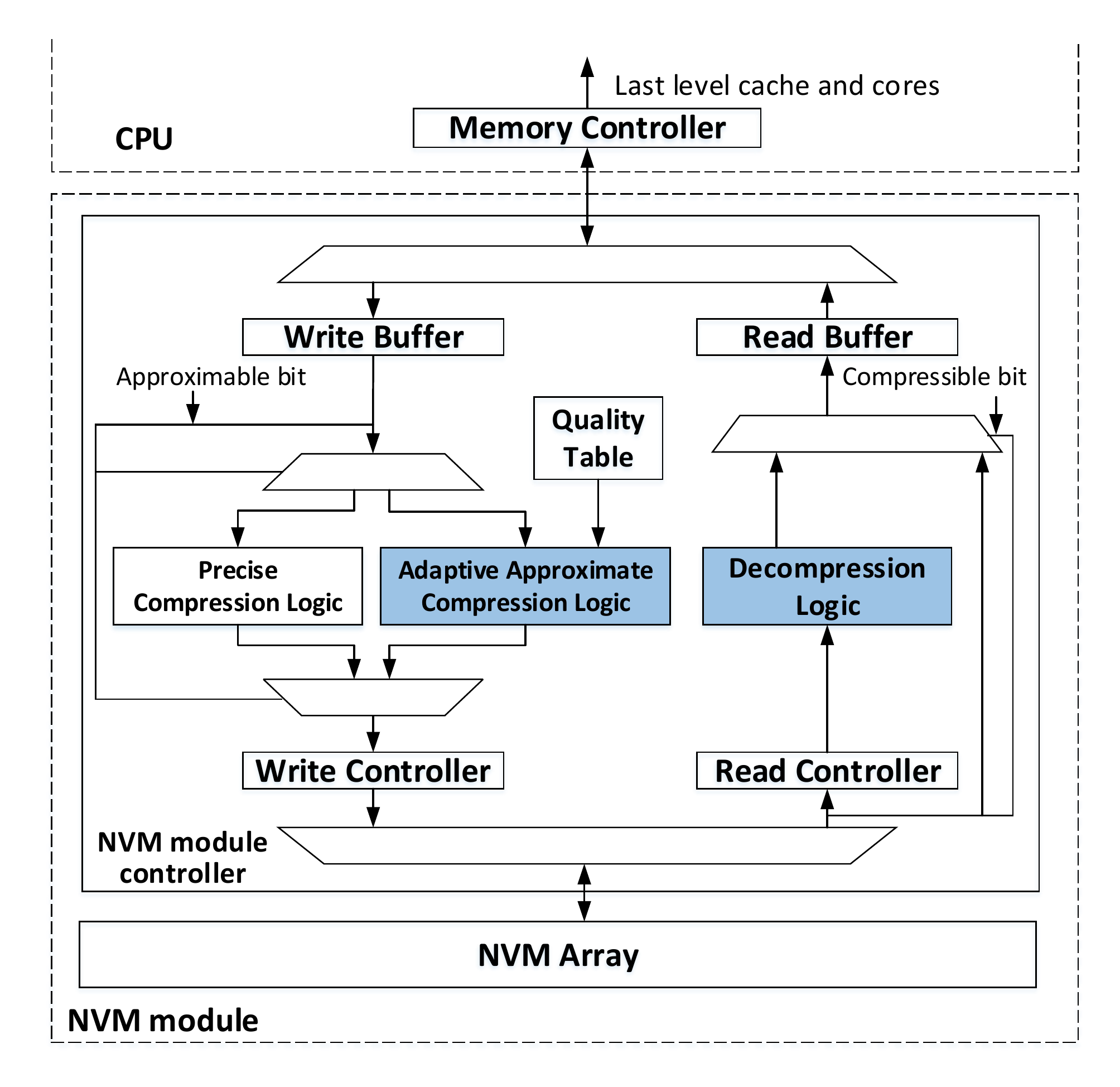}
    \vspace{-0.4cm}
    \caption{\label{fig:Architecture}The architecture overview of SimCom.}
    \vspace{-0.2cm}
\end{figure}

For NVM systems, image-based applications incur many incompressible
writes to NVMs, which result in high write latency and energy consumption.
SimCom achieves high compression performance via the pixel-level
similarity in data writes. The efficient
approximate compression reduces the data size and improves 
the memory performance in NVM systems.

Figure~\ref{fig:Architecture} shows the hardware architecture overview 
of SimCom. Specifically, \emph{Adaptive Approximate Compression Logic} 
and \emph{Decompression Logic} respectively implement the compression and decompression 
schemes of SimCom.
\emph{Quality Table} is an on-chip cache~\cite{DBLP:conf/date/RanjanVPV0R17,DBLP:conf/iccad/ZhaoXCZ17}, 
which stores the start and end addresses of memory regions with 
\emph{Approximation Factor} as \emph{AF} (i.e., approximation degrees) for images.
These addresses and \emph{AF} are specified through programmer 
annotations~\cite{DBLP:conf/pldi/SampsonDFGCG11,sampson2015accept,DBLP:conf/micro/MiguelAMJ15,DBLP:conf/date/RanjanVPV0R17}
and transported to \emph{Quality Table} via ISA extensions~\cite{DBLP:conf/asplos/EsmaeilzadehSCB12}.
An approximable bit in a tag entry indicates the precision of data blocks 
in caches~\cite{DBLP:conf/asplos/EsmaeilzadehSCB12,DBLP:conf/micro/MiguelAMJ15}.

In a write access, we assume approximable data are 
indicated by approximable bits like prior 
works~\cite{DBLP:conf/asplos/EsmaeilzadehSCB12,DBLP:conf/micro/MiguelAMJ15}. 
For approximable data, \emph{Adaptive Approximate Compression Logic} 
finds continuous similar words and compresses them 
into base words and runs. For \emph{precise data}, i.e., data
containing any data except for bitmaps, \emph{Precise
Compression Logic} uses existing precise compression schemes instead 
(e.g., FPC~\cite{DBLP:conf/nanoarch/DgienPHLM14}). In a read access, 
compressed data indicated by compressible bits are decompressed 
in \emph{Decompression Logic} before responding to requests. 
In the decompression stage, if the approximable bit is set to 1, the approximate 
decompression is used; otherwise, the precise decompression is used.

\subsection{Similarity Metric}
\label{sec:SimilarityMetric}

We use normalized difference between two partitioned words to 
quantify the similarity. Only if the normalized difference is smaller 
than \emph{AF}, two words are considered to be similar.
The normalized difference between words $p$ and $q$ is calculated using 
Equation~\ref{eq:NormDiff}. 

\vspace{-0.2cm}
\begin{equation}
    normDiff=\frac{max\{|p[i]-q[i]|\}}{maxValue},i\in[0,CC) \label{eq:NormDiff}
\end{equation}
\vspace{-0.2cm}

The difference between words $p$ and $q$ is normalized to the maximal 
value of per channel for a pixel ($maxValue$), which is determined 
by \emph{BPC}. When \emph{BPC} is 8, $maxValue$ is 255.

\subsection{Uniform Data Partition}
\label{sec:UniformPartition}

In order to efficiently compress the data writes, 
we need to partition the data without 
decreasing the similarity in bitmaps.
The intuitive solution is to partition the data at the pixel 
boundaries whereby the partitioned words would be
similar, since these words correspond to adjacent pixels in bitmaps. 
However, how to identify pixel boundaries in data
becomes a new challenge. 
We can't figure out the positions of pixel boundaries without
additional context information, such as the offset of the data
in bitmaps and the corresponding bitmap format. A straightforward
solution is to allocate each pixel with a fixed alignment 
(the alignment should be a factor of the data write size, e.g., 4 B),
thus enabling fixed pixel boundary positions in data.
However, when the actual pixel size (e.g., 3 B) mismatches the alignment, 
it significantly decreases storage density and wastes storage space.

In order to preserve the similarity in partitioned words with low 
overheads, we propose a uniform scheme to 
partition the data in a write access. Due to the pixel-level similarity,
the data form an approximate periodic cycle of the pixel size.
Therefore, we advocate partitioning 
at the granularity of pixel size and leave the possible remaining bytes 
(when the data size isn't a multiple of the pixel size) at the end 
as a partial word, called \emph{remainder}. For example, if the pixel
size is 3 bytes and the data write size is 64 bytes, data
are divided into 21 words of 3 bytes and 1 partial
word of 1 byte. As shown in Figure~\ref{fig:SimComExample}, though 
the partitioned positions are not pixel boundaries,
the words containing data from adjacent pixels are 
similar due to the periodic cycle. 

\subsection{Search for Continuous Similar Words}

Since continuous similar words require that any two words are 
similar (\secref{sec:IntraBlockSimilarity}), the time 
complexity to get a group of exact continuous similar words is $O(n^2)$ 
($n$ denotes the number of words). The high time complexity incurs high 
latency and hardware overhead to accurately find out all continuous 
similar words in a write access. 

In order to alleviate the cost of searching 
similar words during compression, we propose to approximately search
for continuous similar words. Specifically, we slightly relaxed the 
requirements for each group of continuous similar words. The words in 
relaxed continuous similar words are only required to be similar 
to the base word (for simplicity, we use continuous similar words
to represent relaxed continuous similar words in the following text unless
specified). 
The accuracy of approximate search is still constrained by the \emph{AF}.
The reason is that even with the approximation in search, the 
normalized difference threshold for each group is only \emph{2AF}.

Though the appropriate candidate of base word for a group of similar words is the 
average, we take the first word as the base word for two reasons (in the following text, 
we use base word and base interchangeably): (1) Taking the first word 
as the base simplifies the compression logic; (2) Due to the selection 
of the first word as a base, the compression performance loss is slight.

With the relaxation in similarity and selection of the base for
continuous similar words, the time complexity of getting continuous 
similar words decreases to $O(n)$, which efficiently decreases the
complexity of compression logic and improves the compression 
performance.

\subsection{Compression-aware NVMs}
\label{sec:CompressionWorkflow}

\begin{figure}[!ht]
    \vspace{-0.2cm}
    \centering
    \includegraphics[width=0.46\textwidth]{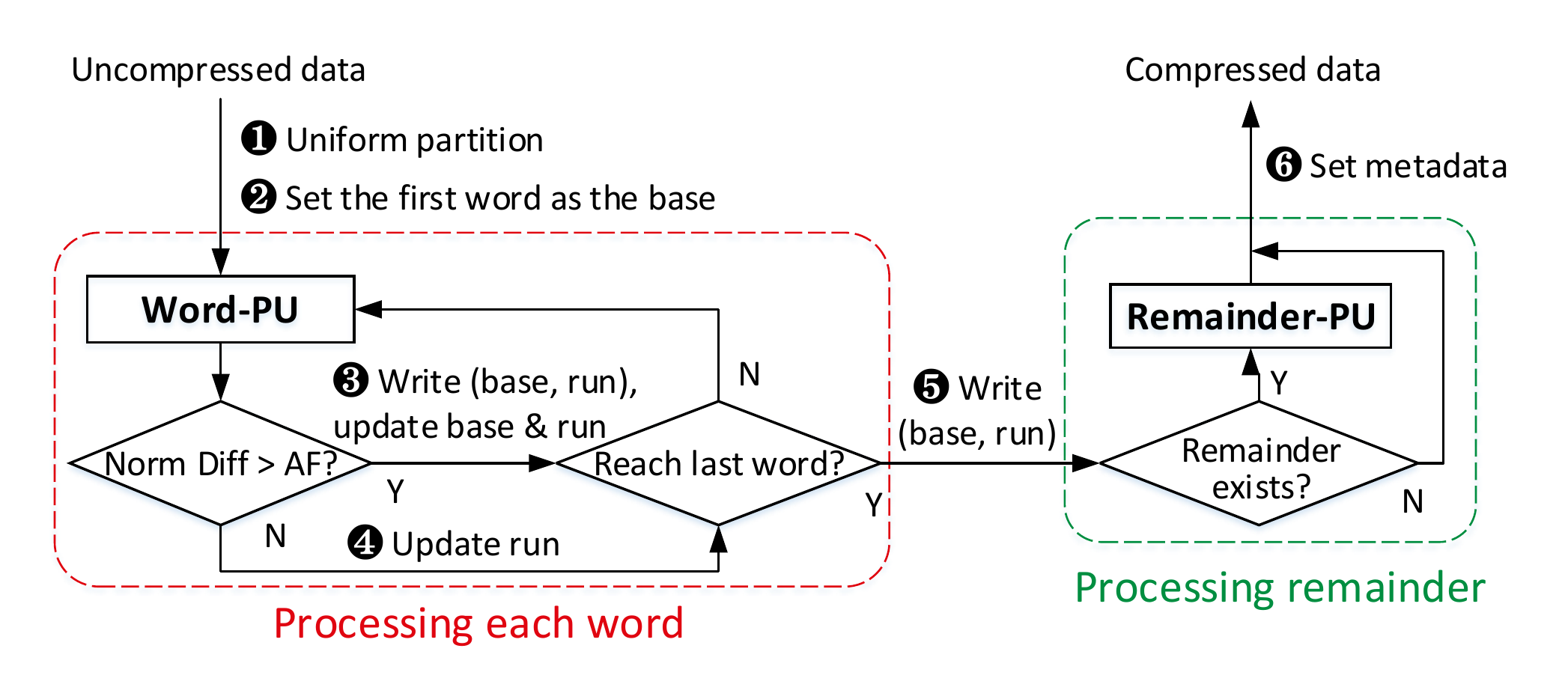}
    \vspace{-0.2cm}
    \caption{\label{fig:ApproximateCompression}The workflow of 
    approximate compression.}
    \vspace{-0.2cm}
\end{figure}


\textbf{Write:}
If the approximable bit of a write access is set to 1, approximate compression is used.
Figure~\ref{fig:ApproximateCompression} illustrates the workflow of approximate 
compression for incoming data. After uniformly partitioning
the data into words and possible remainder \ballnumber{1}, SimCom sets the first
word as the initial base \ballnumber{2}. The Word Processing Unit (\emph{Word-PU})
uses Equation~\ref{eq:NormDiff} to calculate the normalized difference between
the base and rest words. If the normalized difference is larger than the \emph{AF},
current values of base and run are written into compressed data. 
Current word is set as the new base and the run is reset to 0 \ballnumber{3}. 
If the normalized difference is no larger than the threshold, SimCom 
only increases the run by one \ballnumber{4}. After 
processing each word, SimCom records the last pair of base and run 
in the compressed data \ballnumber{5}. When the remainder is available 
in the current partition, the Remainder Processing Unit (\emph{Remainder-PU})
obtains the normalized difference between the remainder and the last 
base using Equation~\ref{eq:NormDiff} with the \emph{CC} substituted by 
the number of channels in the remainder. If the normalized difference 
is larger than the \emph{AF}, SimCom 
writes the remainder and sets the Most Significant Bit (MSB) of the run, 
called \emph{remainder bit}, to indicate the existence of the remainder 
in the compressed data.
SimCom uses the first byte of compressed data to record the number of 
bases \ballnumber{6}. 
If the approximable bit is reset to 0, existing precise compression schemes
(e.g., FPC) are used. For both approximate and precise compression,
if compressed data size is smaller than that of original data, SimCom writes 
compressed data and sets the compressible bit to 1; otherwise, SimCom
writes uncompressed data and resets the compressible bit to 0.

\noindent
\textbf{Read:}
If both the compressible bit and the approximable bit are set to 1, 
approximate decompression is used to reconstruct the stored data.
Specifically, for each pair of base and run in compressed data, 
the base is used to fill in the decompressed data multiple times 
according to the run.
The remainder bit is checked: if set to 1, the remainder in the compressed data is 
used to complement the decompressed data; otherwise, the last base is used. 
If only the compressible bit is 1, compressed data are reconstructed by
the inverse procedure of precise compression (e.g., FPC).
If the compressible bit is 0, the data would bypass the 
decompression procedure and respond to read accesses.

\begin{figure}[!ht]
    \vspace{-0.3cm}
    \centering
    \includegraphics[width=0.46\textwidth]{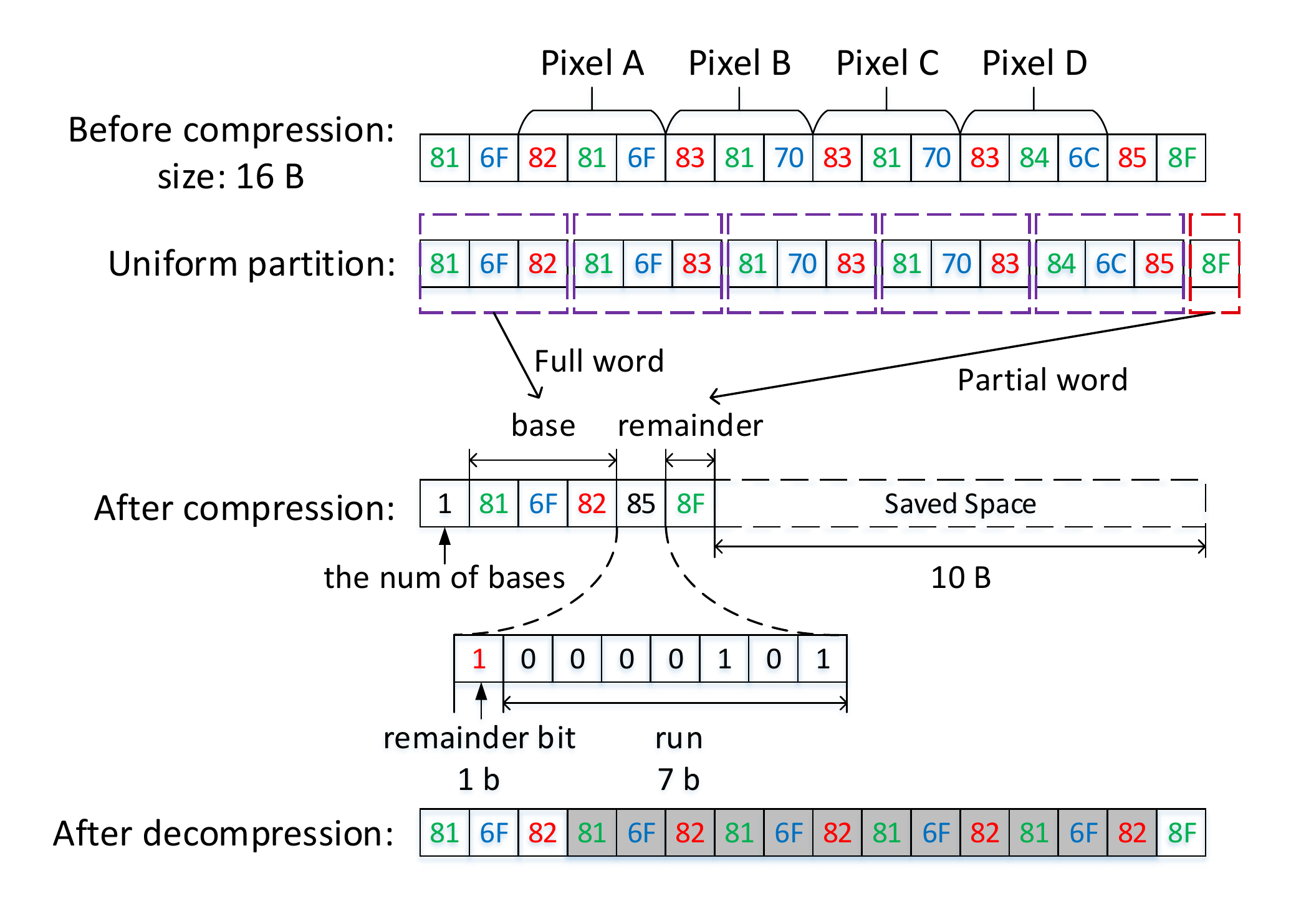}
    \vspace{-0.3cm}
    \caption{\label{fig:SimComExample}An example of approximate compression and 
    decompression scheme. \emph{(AF = 0.05)}}
    \vspace{-0.2cm}
\end{figure}

Figure~\ref{fig:SimComExample} shows an
example of approximate compression/decompression for data writes of 16 bytes. 
Although the partitioned positions aren't boundaries among pixels A, B, C, and D,
partitioned words show pixel-level similarity and form a group of continuous similar words with
a remainder of one byte. By default, the first word in similar words is selected as
the base. In this example, the normalized difference between the remainder and the base
exceeds \emph{AF}. Therefore, the remainder is placed at the end of compressed
data and the remainder bit of last run is 1. After compression, the data size is 
reduced by 10 bytes. During decompression, the base is used to 
fill in similar words (i.e., the shadowed bytes in the figure).

\section{Adaptive Approximate Compression}
\label{sec:AdaptiveCompression}



In order to handle different bitmap formats in the 
compression/decompression logic, the approximate compression 
proposed in \secref{sec:ApproximateCompression} 
requires extra metadata including \emph{CC} and \emph{BPC}.
Though it's possible to annotate the metadata to be stored in cache 
tags~\cite{DBLP:conf/micro/MiguelAMJ15} or address table in memory
controllers~\cite{DBLP:conf/date/RanjanVPV0R17,DBLP:conf/islped/RanjanRRR17}, 
these techniques cause additional overheads and programmer annotations.
Moreover, users need to confirm the bitmap formats and annotate
these metadata before execution. Hence, in this section, 
we propose to leverage the image characteristics and 
adaptively select the appropriate mode for data compression
without additional programmer annotations.

\subsection{Adaptive Compression Scheme}
\label{sec:AdaptiveCompressionScheme}



\textbf{\textit{1) Why use predefined compression modes for different image formats?}}
The images generally include grayscale and color images.
Grayscale images contain only one channel and the color images in RGB color space consist of
red, green, blue, as well as the optional alpha channel. 
In other color spaces (e.g., YUV), similar components (e.g., one luminance channel
and two chrominance channels) exist.
The \emph{BPC} in common 
images is 8 bits representing 256 levels in each channel. 24 bits per pixel ($3C1B$) represent
more than 16 million colors, while the number of colors discriminated by the 
human eye is up to 10 million~\cite{judd1975color,leong2006number}. For applications
processing HD (high-definition) images, 
16 bits per channel is enough to encode the necessary colors. 
Therefore, we propose to use
six compression modes to handle different image formats. 
The options for \emph{CC} are 1 (e.g., grayscale),
3 (e.g., RGB), and 4 (e.g., RGB with the alpha channel) and the options
for \emph{BPC} are 8 and 16.

\begin{figure}[!ht]
    \vspace{-0.3cm}
    \centering
    \includegraphics[width=0.4\textwidth]{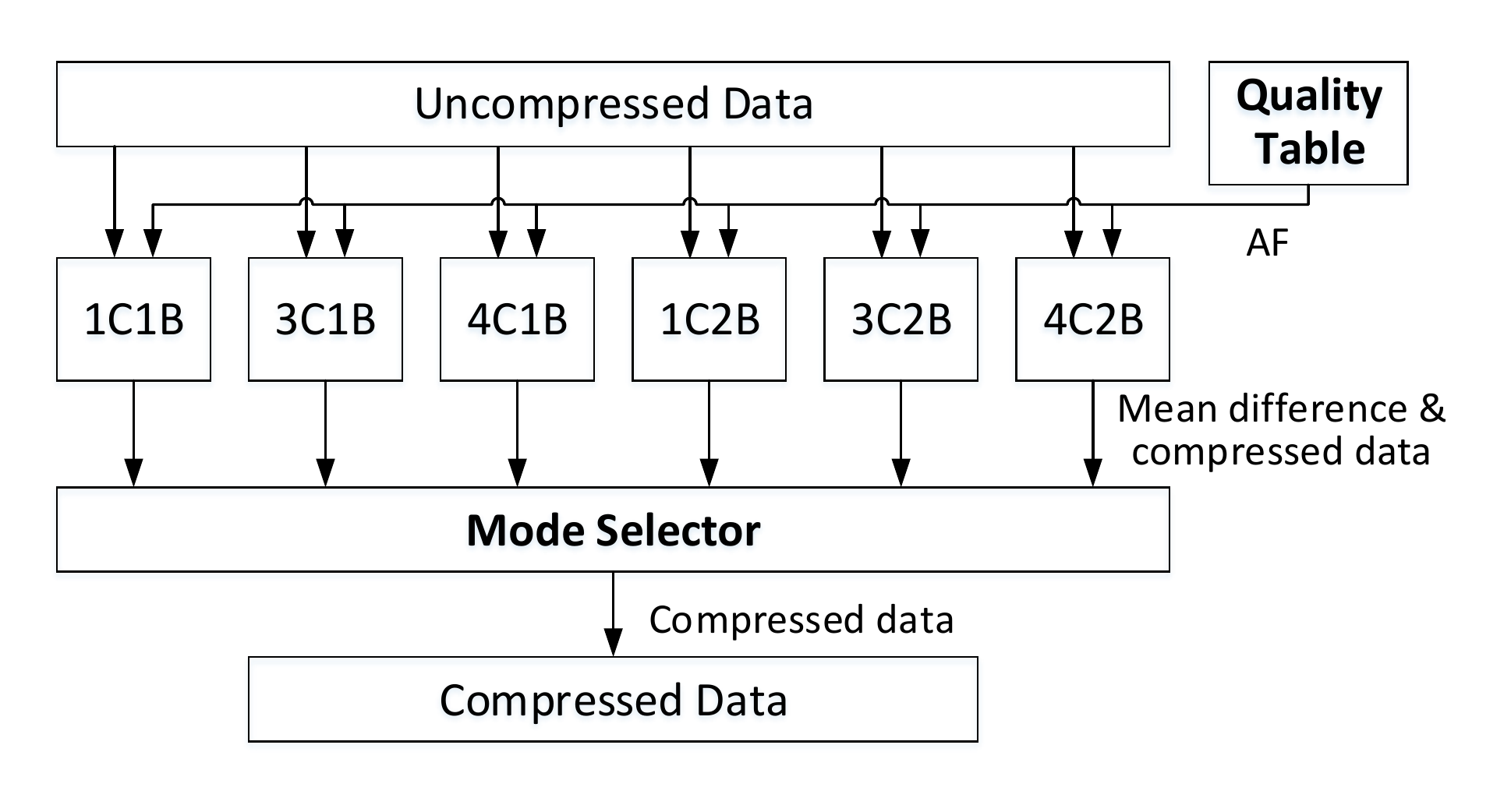}
    \vspace{-0.2cm}
    \caption{\label{fig:AdaptiveCompression}Adaptive 
    compression scheme overview. \emph{(The two integers in each 
    compression mode denote the number of channels and the bytes 
    per channel.)}}
    \vspace{-0.2cm}
\end{figure}

\textbf{\textit{2) How to determine the suitable compression mode for a write access?}}
A straightforward approach is to sample some write accesses
for an efficient compression mode to be applied on later NVM writes.
Sampling works when all write accesses have regular pattern formats
(e.g., all applications using one bitmap format). 
However, sampling often fails when data writes have random pattern formats
(e.g., applications using different bitmap formats 
are running in an NVM system).
Instead of sampling, 
SimCom performs six compression modes in 
parallel and selects the compression mode with the minimal mean difference.
Mean difference indicates the average difference between every two adjacent words in 
data. We observe that the mean difference of the right compression 
mode (i.e., the mode matching the bitmap format) is minimal, which makes sense 
due to the pixel-level similarity. 
Figure~\ref{fig:AdaptiveCompression} shows the overview of adaptive 
compression scheme used in SimCom. Six compression
modes with different \emph{CC} and \emph{BPC} process data in parallel.
\emph{Mode Selector} first selects the mode with the minimal 
mean difference. If multiple modes have the minimal mean 
difference, \emph{Mode Selector} chooses the one with the minimal 
compressed data size. For the simplicity of compression logic, SimCom reuses the
normalized difference between each word and the base as the difference
between adjacent words. Due to the error-tolerance of application and the 
similarity between words and their bases, the reuse of normalized difference is
acceptable. The experimental evaluation in \secref{sec:Evaluation} 
verifies the correctness of mode selection and result reuse.

\subsection{Metadata Management}

There are two classes of metadata in SimCom. The first-class metadata
are used for approximately compressed data in SimCom
including the choice of
compression mode, the number of bases, and one remainder bit.
When using $1C1B$ compression mode, each pair of base and run 
occupy 2 bytes. Therefore, there are no more than 32 pairs of base 
and run in compressed data with 64-byte write data granularity. 
For other compression modes, the number of bases in the compressed data
is smaller. As a result, we only use 5 bits to record 
the number of bases. SimCom uses the first byte of the compressed data as 
metadata. The highest 3 bits of metadata are used to encode the 
choice of compression mode and the rest 5 bits for the number of 
bases in the compressed data. 
As described in \secref{sec:CompressionWorkflow}, the MSB of the last
run is used as the remainder bit to indicate if a remainder exists in 
the compressed data. Hence, the first-class metadata are 
stored in compressed data.
The second-class metadata are used for approximate
compression including one approximable bit and one compressible bit
(\secref{sec:DesignOverview}). SimCom stores the approximable
and compressible bits in a separate region in NVMs like prior
works~\cite{DBLP:conf/IEEEpact/PekhimenkoSMGKM12,
DBLP:conf/nanoarch/DgienPHLM14,DBLP:conf/micro/MiguelAMJ15,
DBLP:conf/islped/RanjanRRR17}. However, the two bits can be packed
into compressed data to reduce NVM accesses and improve memory 
bandwidth like recent works~\cite{DBLP:conf/micro/HongNABKH18,
DBLP:journals/corr/abs-1807-07685}.

\section{Performance Evaluation}
\label{sec:Evaluation}

\subsection{Experimental Setting}
\label{sec:ExperimentalSetting}

\begin{table}[!ht]
  \centering
  \caption{System configurations.}
  \vspace{-0.2cm}
  \label{tab:SystemConfig}
  \scalebox{0.85}{
    \begin{tabular}{|c|M{5.2cm}|}
      \hline
      \multicolumn{2}{|c|}{\textbf{Processor}} \\
      \hline
        CPU & 1 core, X86-64 processor, 2 GHz\\ \hline
        L1 I/D cache & 32 KB, 2 ways, LRU\\ \hline
        L2 cache & 1024 KB, 8 ways, LRU\\ \hline
        Cache block size & 64 B\\ \hline
      \multicolumn{2}{|c|}{\textbf{Main Memory using PCM}} \\ \hline
        Memory controller & FCFRFS\\ \hline
        Read/Write latency & 120 ns/150 ns\\ \hline
        Memory organization & 4 GB, 8 B write unit size\\
      \hline
    \end{tabular}
  }
  \vspace{-0.3cm}
\end{table}

We implement SimCom in GEM5~\cite{DBLP:journals/sigarch/BinkertBBRSBHHKSSSSVHW11}
with NVMain~\cite{DBLP:journals/cal/PorembaZ015}. The system configurations of
GEM5 and NVMain are listed in Table~\ref{tab:SystemConfig}. 
Since SimCom focuses on data compression and is orthogonal to the underlying
memory model, we simply use a First Ready First Come First Serve (FRFCFS) memory 
controller to serve NVM accesses. We evaluate the performance with six 
image-based workloads, i.e., jpeg, sobel, 
and kmeans from AxBench~\cite{DBLP:journals/dt/YazdanbakhshMEL17}
and 2dconv, debayer, and histeq from PERFECT~\cite{barker2013perfect}. 
These workloads are selected for various domains, jpeg for compression,
sobel, 2dconv, debayer, and histeq for image processing, and kmeans 
for machine learning.
The ratios of approximable data in 
these workloads are shown in Table~\ref{tab:BitmapBlockRatio}. 
As suggested in~\cite{DBLP:journals/dt/YazdanbakhshMEL17}, we use
RMSE (\secref{sec:BitmapBackground}) as the metric to measure the
output error (i.e., the quality of output image) compared with the
precise compression result. 
The input images come from Kodak dataset~\cite{kodakimage}. 
The output errors are reported using the 
average RMSE of 6 images. Before running these workloads, we warm up
the system with 100 million instructions.

\begin{figure*}[!ht]
  \centering
  \includegraphics[width=.91\linewidth]{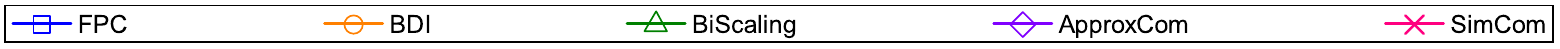}
  \begin{subfigure}{.3\linewidth}
    \includegraphics[width=\linewidth]{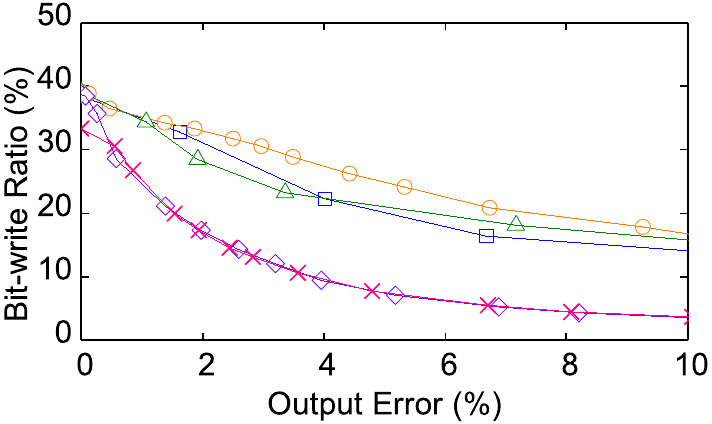}
    \caption{\label{fig:WBRCurveJpeg}Bit-write Ratio}
  \end{subfigure}
  ~ 
  \begin{subfigure}{.3\linewidth}
    \includegraphics[width=\linewidth]{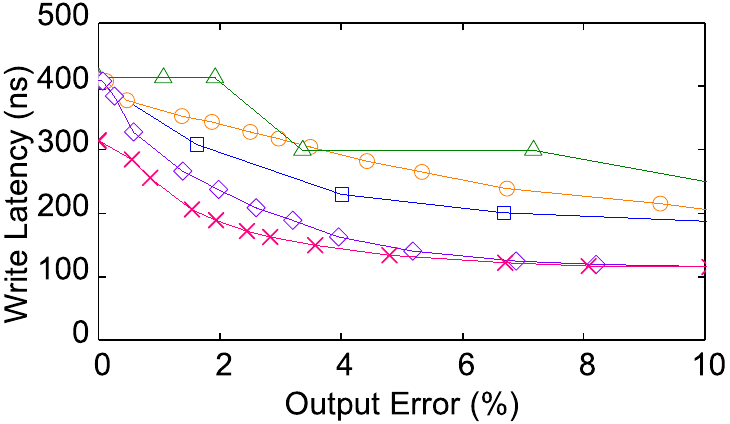}
    \caption{\label{fig:WLCurveJpeg}Write Latency}
  \end{subfigure}
  ~ 
  \begin{subfigure}{.3\linewidth}
    \includegraphics[width=\linewidth]{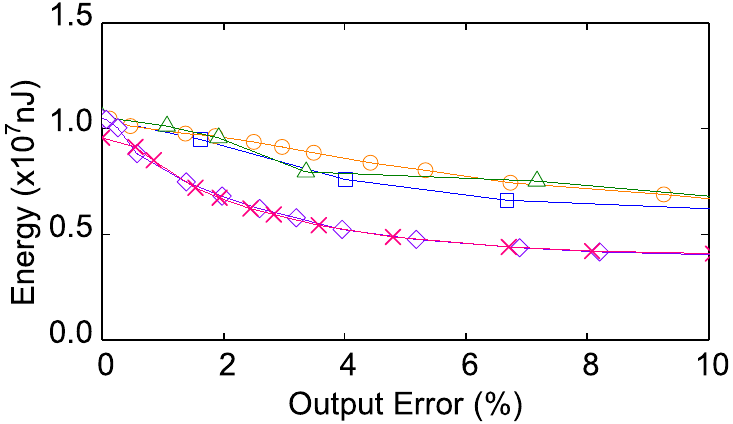}
    \caption{\label{fig:ECurveJpeg}Total Energy Consumption}
  \end{subfigure}
  \vspace{-0.3cm}
  \caption{\label{fig:TradeOffJpeg}The performance of jpeg:
  bit-write ratio, write latency, energy consumption with various output errors.}
  \vspace{-0.2cm}
\end{figure*}

\begin{figure*}[!ht]
  \vspace{-0.2cm}
  \centering
  \includegraphics[width=.91\linewidth]{legend.pdf}
  \begin{subfigure}{.3\linewidth}
    \includegraphics[width=\linewidth]{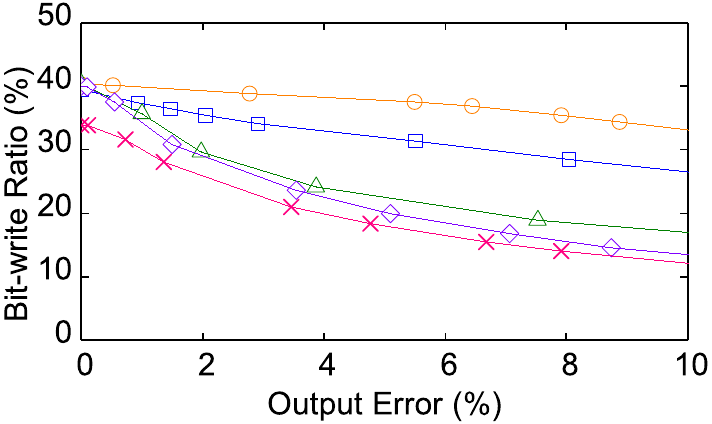}
    \caption{\label{fig:WBRCurveSobel}Bit-write Ratio}
  \end{subfigure}
  ~ 
  \begin{subfigure}{.3\linewidth}
    \includegraphics[width=\linewidth]{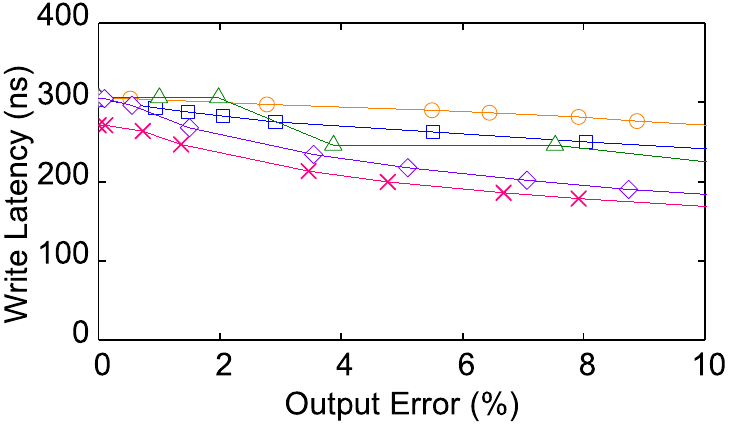}
    \caption{\label{fig:WLCurveSobel}Write Latency}
  \end{subfigure}
  ~ 
  \begin{subfigure}{.3\linewidth}
    \includegraphics[width=\linewidth]{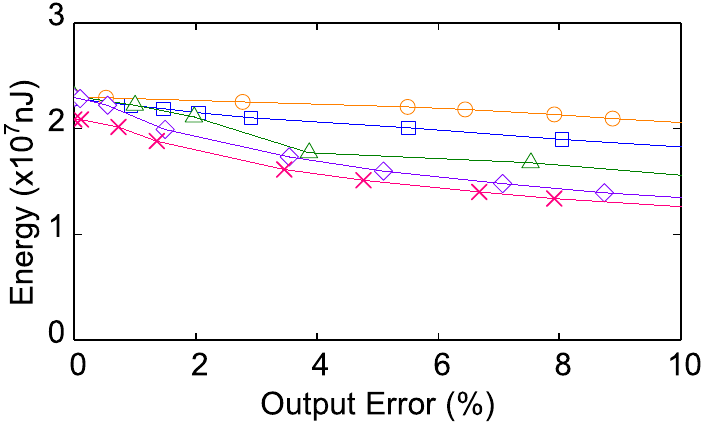}
    \caption{\label{fig:ECurveSobel}Total Energy Consumption}
  \end{subfigure}
  \vspace{-0.3cm}
  \caption{\label{fig:TradeOffSobel}The performance of sobel:
  bit-write ratio, write latency, energy consumption with various output errors.}
  \vspace{-0.2cm}
\end{figure*}

We have evaluated the following compression schemes 
(FNW~\cite{DBLP:conf/micro/ChoL09} is used to further reduce bit-flips 
in all schemes):
\begin{itemize}[leftmargin=*]
  \item \textbf{FPC}: FPC~\cite{DBLP:conf/nanoarch/DgienPHLM14} 
  exploits the general frequent patterns and compresses the matched words 
  with short prefix bits. For fair comparisons, we enhance this scheme by 
  adding approximation. Specifically, when the difference between a
  partitioned word and a relaxed word matching a frequent pattern
  is within the error constraint, the pattern is used to compress
  the word.

  \item \textbf{BDI}: BDI~\cite{DBLP:conf/IEEEpact/PekhimenkoSMGKM12} 
  leverages the narrow value characteristics of array and compresses cache 
  block data into bases with small deltas. This scheme is an approximate version
  of BDI~\cite{DBLP:conf/IEEEpact/PekhimenkoSMGKM12}. It relaxes narrow value
  constraint and compresses the words that slightly overflow the delta limit.

  \item \textbf{BiScaling}: This scheme uses bidirectional precision
  scaling to approximately compress the data to be written~\cite{DBLP:conf/islped/RanjanRRR17}.

  \item \textbf{ApproxCom}: This is our proposed scheme that leverages the
  pixel-level similarity for approximate compression.

  \item \textbf{SimCom}: This is our proposed scheme leveraging 
  pixel-level similarity and adaptive compression (i.e., ApproxCom
  + adaptive compression), which eliminates the annotations on 
  data formats used in BiScaling and ApproxCom.
\end{itemize}

\noindent
Since BiScaling, ApproxCom, and SimCom focus on approximate 
compression on approximable data, we use precise FPC to compress 
precise data in these schemes. 

\renewcommand{\arraystretch}{1.1}
\begin{table}[!ht]
  \centering
  \caption{The annotation requirements in compression 
  schemes. \emph{(\cmark: require the annotation; \xmark: 
  no requirements for the annotation.)}}
  \vspace{-0.2cm}
  \label{tab:Annotations}
  \scalebox{0.7}{
    \begin{tabular}{|c|M{3.4cm}|M{2.3cm}|M{2.6cm}|}
      \hline
      \textbf{Schemes} & \textbf{Approximation Factor} (AF) & \textbf{Channel Count} (CC) & \textbf{Bits Per Channel} (BPC) \\ \hline
      \textbf{FPC} & \cmark & \xmark & \xmark \\ \hline
      \textbf{BDI} & \cmark & \xmark & \xmark \\ \hline
      \textbf{BiScaling} & \cmark & \xmark & \cmark \\ \hline
      \textbf{ApproxCom} & \cmark & \cmark & \cmark \\ \hline
      \textbf{SimCom} & \cmark & \xmark & \xmark \\
      \hline
    \end{tabular}
  }
  \vspace{-0.2cm}
\end{table}
\renewcommand{\arraystretch}{1}

\begin{figure*}[!ht]
  \centering
  \includegraphics[width=.91\linewidth]{legend.pdf}
  \begin{subfigure}{.3\linewidth}
    \includegraphics[width=\linewidth]{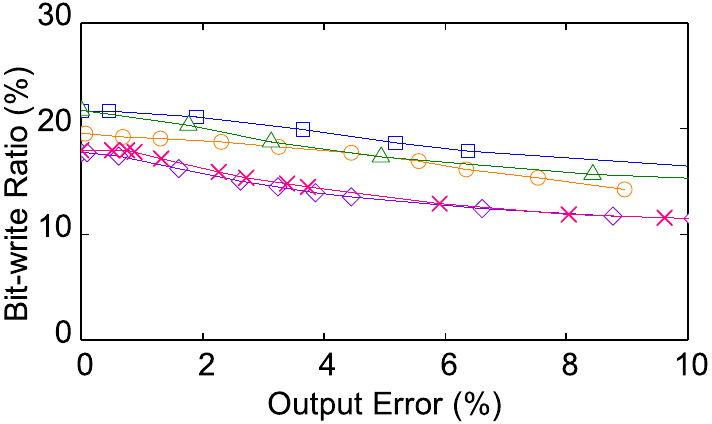}
    \caption{\label{fig:WBRCurveKmeans}Bit-write Ratio}
  \end{subfigure}
  ~ 
  \begin{subfigure}{.3\linewidth}
    \includegraphics[width=\linewidth]{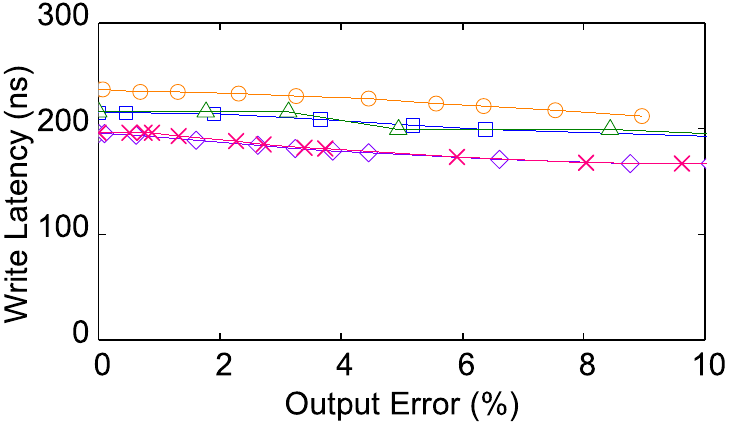}
    \caption{\label{fig:WLCurveKmeans}Write Latency}
  \end{subfigure}
  ~ 
  \begin{subfigure}{.3\linewidth}
    \includegraphics[width=\linewidth]{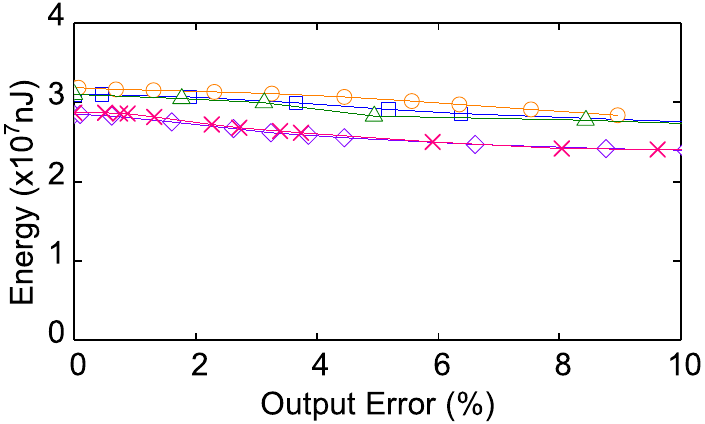}
    \caption{\label{fig:ECurveKmeans}Total Energy Consumption}
  \end{subfigure}
  \vspace{-0.3cm}
  \caption{\label{fig:TradeOffKmeans}The performance of kmeans:
  bit-write ratio, write latency, energy consumption with various output errors.}
  \vspace{-0.2cm}
\end{figure*}

\begin{figure*}[!ht]
  \vspace{-0.2cm}
  \centering
  \includegraphics[width=.91\linewidth]{legend.pdf}
  \begin{subfigure}{.3\linewidth}
    \includegraphics[width=\linewidth]{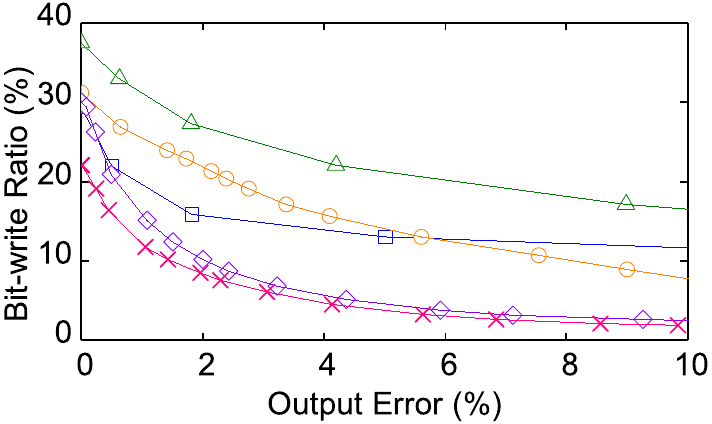}
    \caption{\label{fig:WBRCurve2dconv}Bit-write Ratio}
  \end{subfigure}
  ~ 
  \begin{subfigure}{.3\linewidth}
    \includegraphics[width=\linewidth]{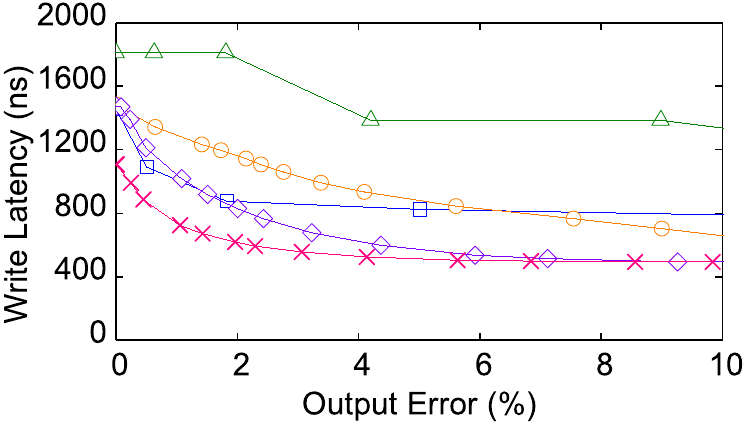}
    \caption{\label{fig:WLCurve2dconv}Write Latency}
  \end{subfigure}
  ~ 
  \begin{subfigure}{.3\linewidth}
    \includegraphics[width=\linewidth]{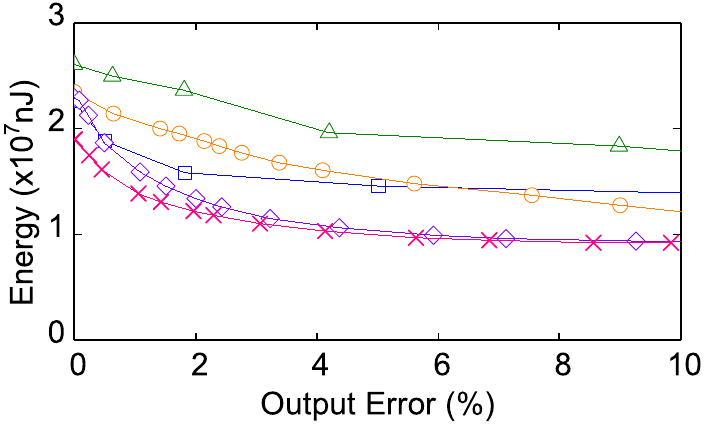}
    \caption{\label{fig:ECurve2dconv}Total Energy Consumption}
  \end{subfigure}
  \vspace{-0.3cm}
  \caption{\label{fig:TradeOff2dconv}The performance of 2dconv:
  bit-write ratio, write latency, energy consumption with various output errors.}
  \vspace{-0.2cm}
\end{figure*}

We leverage programmer 
annotations~\cite{DBLP:conf/pldi/SampsonDFGCG11,sampson2015accept} 
and ISA extensions~\cite{DBLP:conf/asplos/EsmaeilzadehSCB12} to deliver 
necessary information into storage systems like prior 
works~\cite{DBLP:conf/asplos/EsmaeilzadehSCB12,DBLP:conf/micro/MiguelAMJ15,
DBLP:conf/micro/MiguelAJJ16,DBLP:conf/date/RanjanVPV0R17}. 
Programmer annotations are mature techniques and widely used in approximate 
storage systems~\cite{DBLP:conf/pldi/SampsonDFGCG11,
DBLP:conf/asplos/EsmaeilzadehSCB12,DBLP:conf/micro/MiguelAMJ15,
DBLP:conf/micro/MiguelAJJ16,DBLP:conf/date/RanjanVPV0R17}.
We use programmer annotations to annotate bitmaps as approximable data in 
workloads. Through ISA extensions, write accesses with approximable data 
are identified and processed by approximate compression logics.
Table~\ref{tab:Annotations} shows the required annotations for 
all compression schemes. 

\begin{figure*}[!ht]
  \centering
  \includegraphics[width=.91\linewidth]{legend.pdf}
  \begin{subfigure}{.3\linewidth}
    \includegraphics[width=\linewidth]{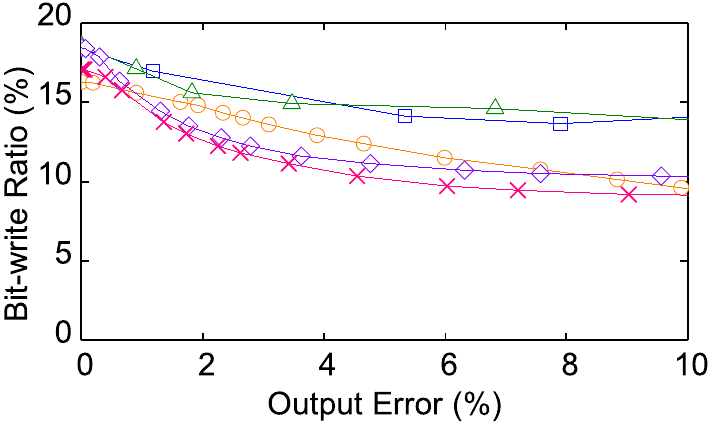}
    \caption{\label{fig:WBRCurveDebayer}Bit-write Ratio}
  \end{subfigure}
  ~ 
  \begin{subfigure}{.3\linewidth}
    \includegraphics[width=\linewidth]{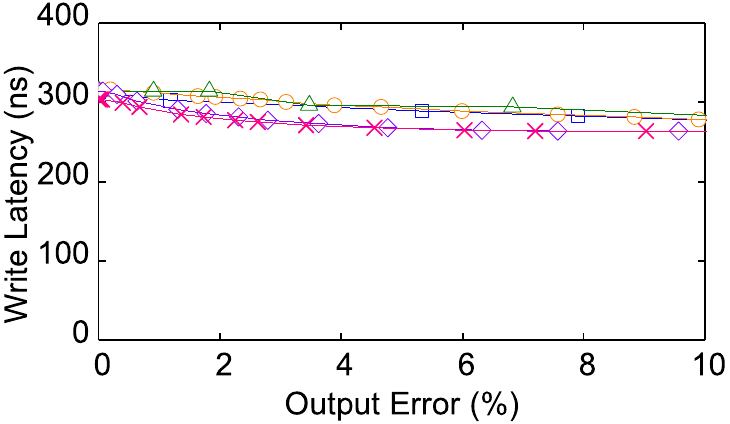}
    \caption{\label{fig:WLCurveDebayer}Write Latency}
  \end{subfigure}
  ~ 
  \begin{subfigure}{.3\linewidth}
    \includegraphics[width=\linewidth]{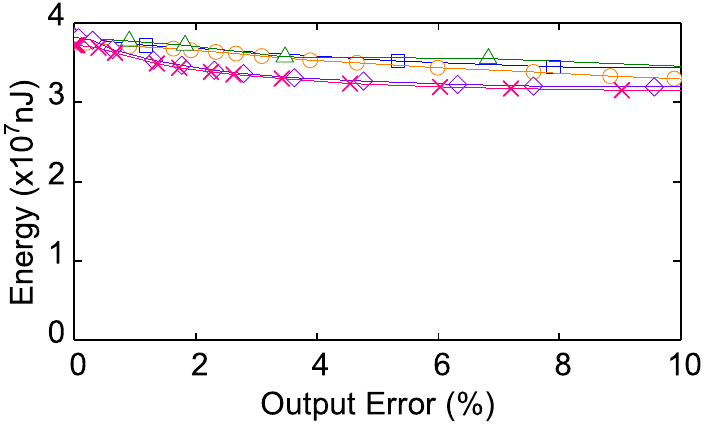}
    \caption{\label{fig:ECurveDebayer}Total Energy Consumption}
  \end{subfigure}
  \vspace{-0.3cm}
  \caption{\label{fig:TradeOffDebayer}The performance of debayer:
  bit-write ratio, write latency, energy consumption with various output errors.}
  \vspace{-0.2cm}
\end{figure*}

\begin{figure*}[!ht]
  \vspace{-0.2cm}
  \centering
  \includegraphics[width=.91\linewidth]{legend.pdf}
  \begin{subfigure}{.3\linewidth}
    \includegraphics[width=\linewidth]{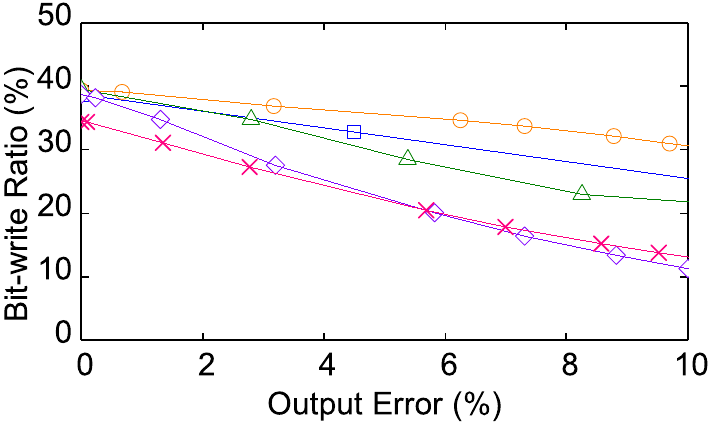}
    \caption{\label{fig:WBRCurveHisteq}Bit-write Ratio}
  \end{subfigure}
  ~ 
  \begin{subfigure}{.3\linewidth}
    \includegraphics[width=\linewidth]{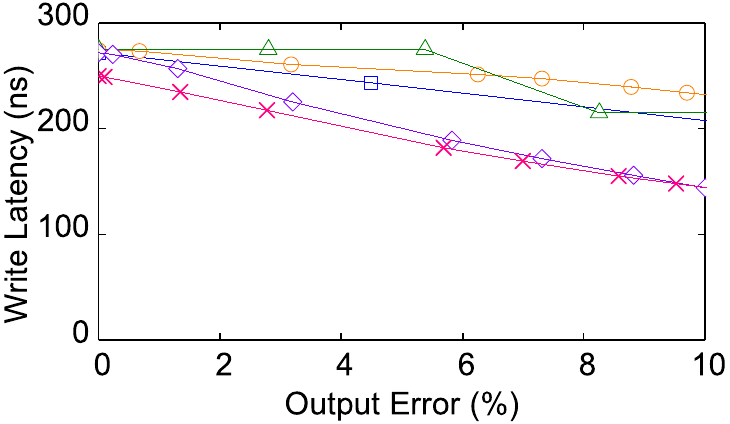}
    \caption{\label{fig:WLCurveHisteq}Write Latency}
  \end{subfigure}
  ~ 
  \begin{subfigure}{.3\linewidth}
    \includegraphics[width=\linewidth]{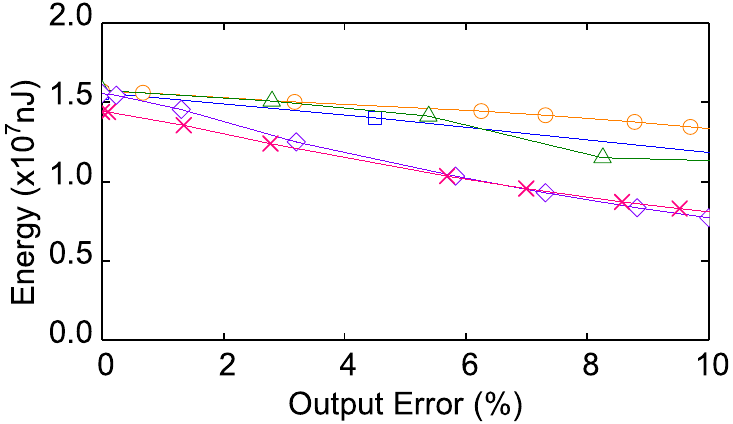}
    \caption{\label{fig:ECurveHisteq}Total Energy Consumption}
  \end{subfigure}
  \vspace{-0.3cm}
  \caption{\label{fig:TradeOffHisteq}The performance of histeq:
  bit-write ratio, write latency, energy consumption with various output errors.}
  \vspace{-0.2cm}
\end{figure*}

\subsection{Quality Efficiency}
\label{sec:EvaluationTradeoff}

Figures \ref{fig:TradeOffJpeg}-\ref{fig:TradeOffHisteq} show the 
the memory performance improvement
in terms of bit-write reduction, write latency, and energy consumption 
with various output errors. 
During our experiments, we observe serious quality degeneration when 
the output error approaches to 10\%. Therefore, we only plot the curves
with output errors under 10\%. 

\textbf{Bit-write Reduction.} 
Figures \ref{fig:WBRCurveJpeg}-\ref{fig:WBRCurveHisteq} show the 
bit-write reduction with different output errors. 
The bit-write ratio denotes the percentage of bit-write after
compression and FNW. A lower bit-write ratio implies a higher
NVM performance improvement.
With the increase of output errors, the bit-write ratios in all approximate
compression schemes decrease. Due to the efficiency of pixel-level similarity,
ApproxCom and SimCom generate less bit-writes than other approximate 
compression schemes with most output errors. SimCom achieves 
8.6\%/10.3\%/8.3\% lower bit-write ratios on average than FPC/BDI/BiScaling 
with the same output error of 3\% (the same constraint
following~\cite{DBLP:conf/islped/RanjanRRR17}). When the output error increases
to 5\%, the average decreases of bit-write ratios become 9.2\%/11.1\%/8.5\%. 
In kmeans and debayer, the benefits of approximation decrease due 
to the smaller ratios of approximable data than other workloads, 
as shown in Table~\ref{tab:BitmapBlockRatio}. Due to the flexibility 
of adaptive compression,
SimCom obtains slightly lower bit-write ratios than ApproxCom.

\textbf{Write Latency.} Figures~\ref{fig:WLCurveJpeg}-\ref{fig:WLCurveHisteq} 
show the write latency with different output errors. Due to the 
electric current constraint in NVMs, the write operation
is divided into several serial write 
units~\cite{DBLP:conf/micro/ChoL09,DBLP:conf/hpca/YueZ13}. Therefore, 
the write latency mainly depends on the data size.
Compared with precise FPC and precise BDI, SimCom gains average write 
latency reduction of 33.0\% and 34.8\% subject to 3\% output error. 
When the constraint of quality loss is relaxed to 5\%, the average 
reduction of write latency become 38.2\% and 40.0\%.
For approximate compression schemes, the superiority of SimCom in terms of bit-write ratio 
turns into the benefits in write latency. Under 3\% and 5\% quality loss, SimCom 
achieves average 21.5\%/28.2\%/30.3\% and 24.0\%/30.1\%/31.6\% write 
latency reduction compared with FPC/BDI/BiScaling, respectively.

\textbf{Energy Consumption.} 
Figures~\ref{fig:ECurveJpeg}-\ref{fig:ECurveHisteq} show the
energy consumption with various quality loss. 
Since the energy consumed in the programming process is the
main fraction in total energy consumption of NVMs~\cite{DBLP:conf/hpca/PalangappaM16},
the number of bit-writes determines the energy consumption. 
By decreasing only 3\% output quality,
SimCom obtains 28.3\% and 29.0\% energy savings than precise FPC and precise BDI.
The average energy savings become 34.7\% and 35.2\% when the quality loss constraint 
is 5\%. Compared with approximate
compression schemes, SimCom reduces the consumed energy by 19.1\%/23.0\%/21.6\% and
22.4\%/26.3\%/24.0\% than FPC/BDI/BiScaling with 3\% and 5\% quality loss, respectively.


\vspace{-0.2cm}
\subsection{Breakdown of Bit-write Reduction}

\begin{figure}[!ht]
  \begin{subfigure}{\linewidth}
    \centering
    \includegraphics[width=.95\linewidth]{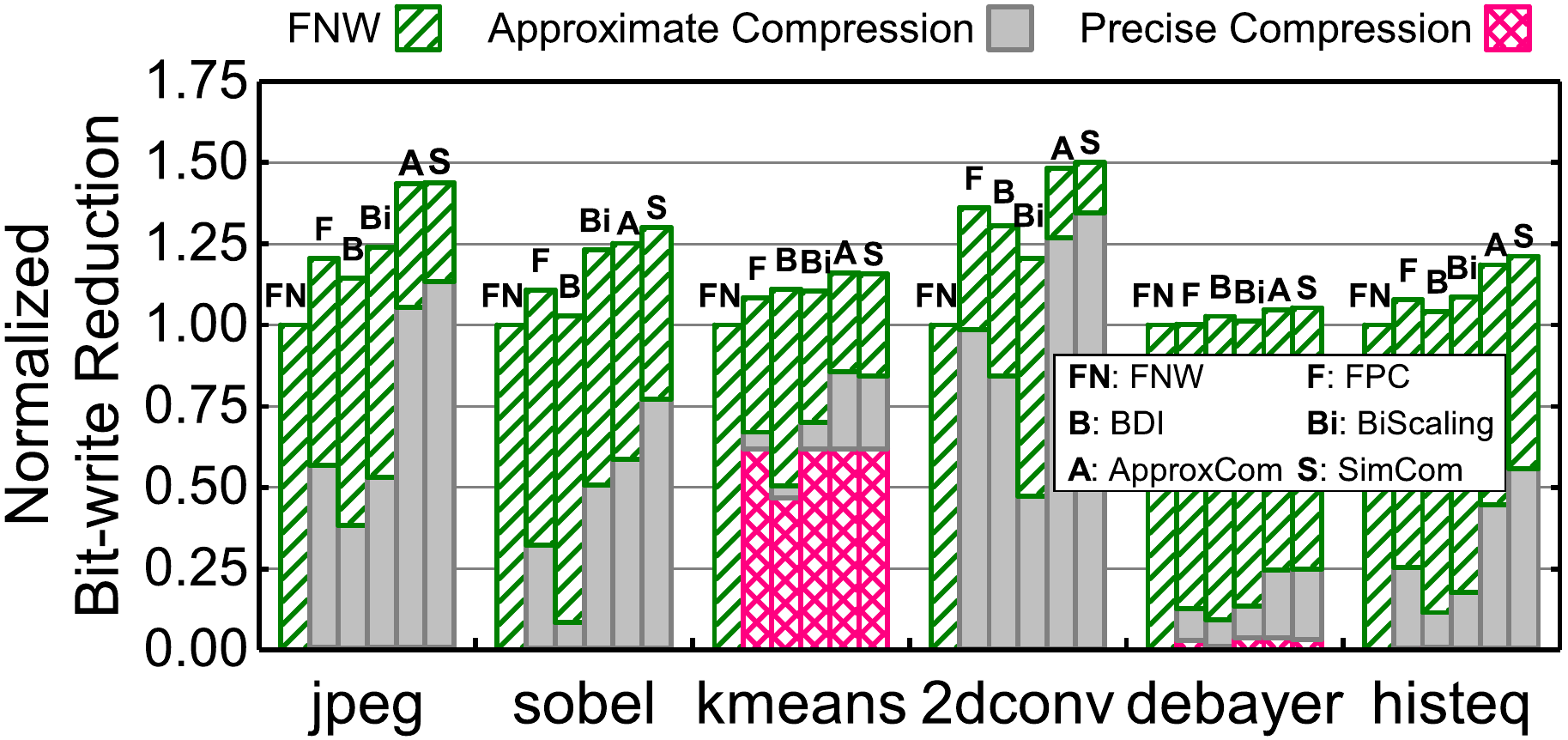}
    \caption{\label{fig:WBRContributionE3}output error \textless{} 3\%}
  \end{subfigure}\par\medskip
  \begin{subfigure}{\linewidth}
    \centering
    \includegraphics[width=.95\linewidth]{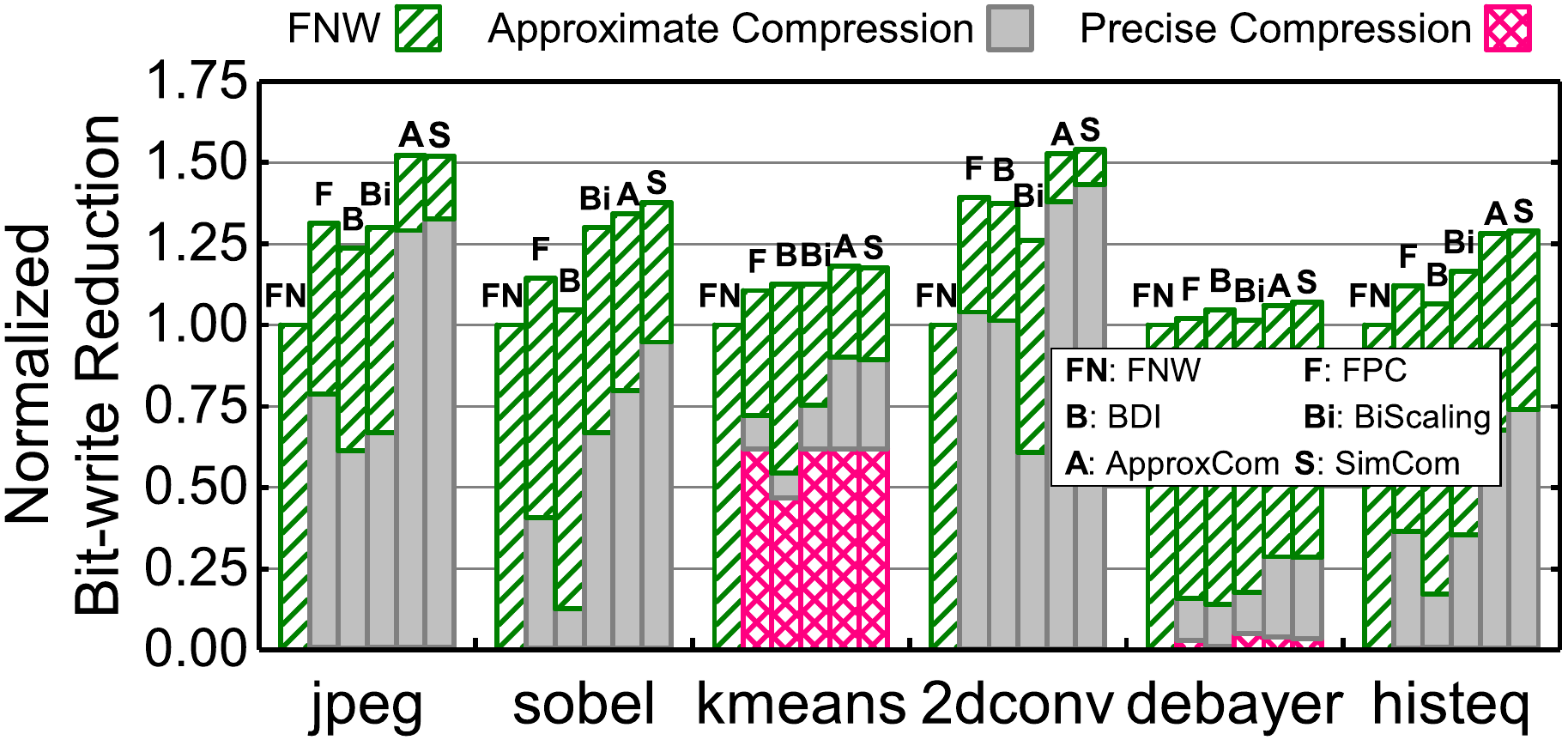}
    \caption{\label{fig:WBRContributionE5}output error \textless{} 5\%}
  \end{subfigure}
  \caption{\label{fig:WBRContribution}Breakdown of bit-writes reduction with various output error.}
\end{figure}

In order to evaluate the contribution of different techniques (i.e., precise compression 
for precise data, approximate compression for approximable data, and FNW for 
compressed data) in all evaluated 
schemes, we record the bit-write reduction from each technique and present
the results in Figures~\ref{fig:WBRContributionE3}-\ref{fig:WBRContributionE5}. 
We set the output error constraints as 3\%~\cite{DBLP:conf/islped/RanjanRRR17} 
and 5\% (aggressive approximation) for evaluation.

As shown in 
Figures~\ref{fig:WBRContributionE3}-\ref{fig:WBRContributionE5}, SimCom gains the largest
bit-write reduction from the approximate compression for error-tolerant data,
thus obtaining the advantage of bit-write reduction over other schemes.
Though the percentages of precise data are large in debayer and histeq,
the precise compression performance is poor due to the pattern mismatch and 
irregular data types in these workloads, thus resulting in the inefficiency 
of precise compression (i.e., FPC and BDI). 

\subsection{Output Quality}

\begin{figure}[!ht]
  \centering
  \begin{subfigure}{.44\linewidth}
    \includegraphics[width=\linewidth]{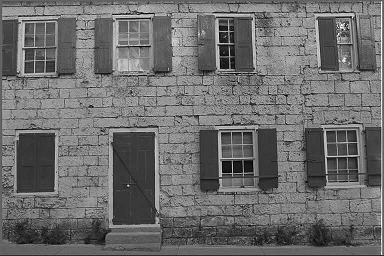}
    \caption{\label{fig:JpegE0}original output}
  \end{subfigure}
  ~ 
  \begin{subfigure}{.44\linewidth}
    \includegraphics[width=\linewidth]{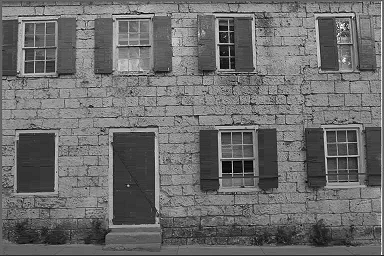}
    \caption{\label{fig:JpegE3}output error \textless{} 3\%}
  \end{subfigure}
  \vspace{-0.3cm}
  \caption{\label{fig:JpegOutput}Output quality of jpeg.}
  \vspace{-0.2cm}
\end{figure}
\begin{figure}[!ht]
  \vspace{-0.2cm}
  \centering
  \begin{subfigure}{.44\linewidth}
    \includegraphics[width=\linewidth]{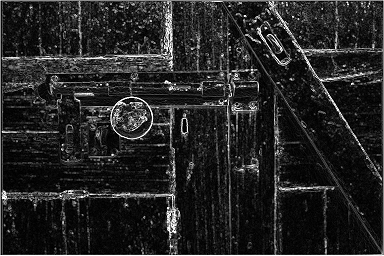}
    \caption{\label{fig:SobelE0}original output}
  \end{subfigure}
  ~ 
  \begin{subfigure}{.44\linewidth}
    \includegraphics[width=\linewidth]{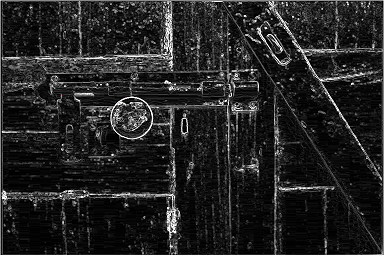}
    \caption{\label{fig:SobelE3}output error \textless{} 3\%}
  \end{subfigure}
  \vspace{-0.3cm}
  \caption{\label{fig:SobelOutput}Output quality of sobel.}
  \vspace{-0.2cm}
\end{figure}
\begin{figure}[!ht]
  \vspace{-0.2cm}
  \centering
  \begin{subfigure}{.44\linewidth}
    \includegraphics[width=\linewidth]{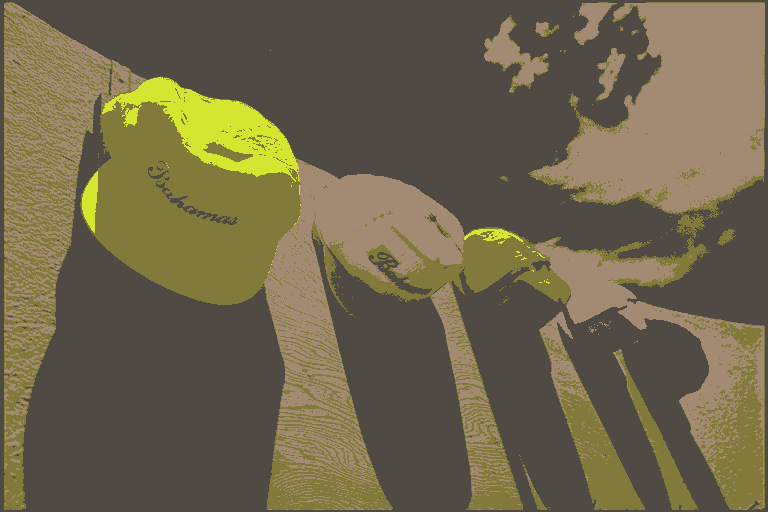}
    \caption{\label{fig:KmeansE0}original output}
  \end{subfigure}
  ~ 
  \begin{subfigure}{.44\linewidth}
    \includegraphics[width=\linewidth]{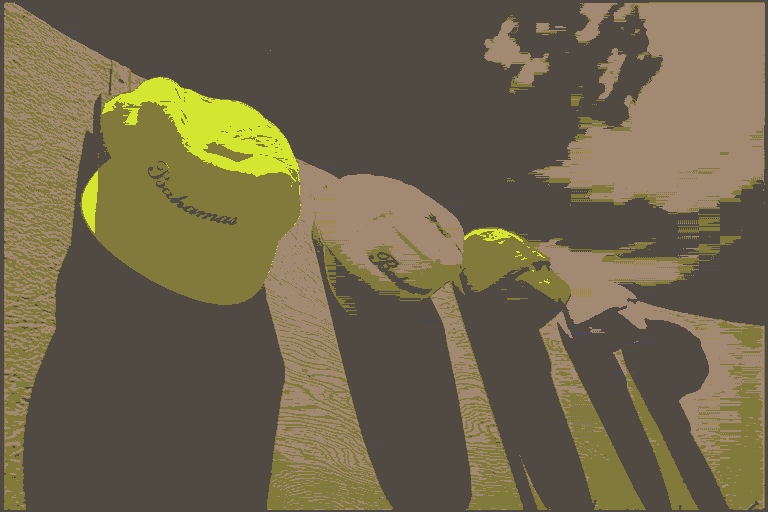}
    \caption{\label{fig:KmeansE3}output error \textless{} 3\%}
  \end{subfigure}
  \vspace{-0.3cm}
  \caption{\label{fig:KmeansOutput}Output quality of kmeans.}
  \vspace{-0.2cm}
\end{figure}
\begin{figure}[!ht]
  \vspace{-0.2cm}
  \centering
  \begin{subfigure}{.44\linewidth}
    \includegraphics[width=\linewidth]{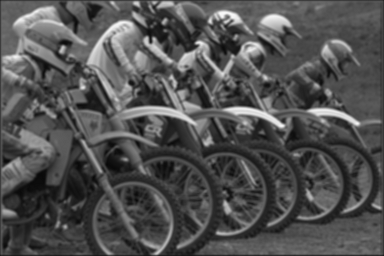}
    \caption{\label{fig:2dconvE0}original output}
  \end{subfigure}
  ~ 
  \begin{subfigure}{.44\linewidth}
    \includegraphics[width=\linewidth]{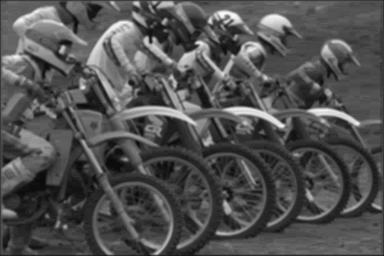}
    \caption{\label{fig:2dconvE3}output error \textless{} 3\%}
  \end{subfigure}
  \vspace{-0.3cm}
  \caption{\label{fig:2dconvOutput}Output quality of 2dconv.}
  \vspace{-0.2cm}
\end{figure}
\begin{figure}[!ht]
  \vspace{-0.2cm}
  \centering
  \begin{subfigure}{.44\linewidth}
    \includegraphics[width=\linewidth]{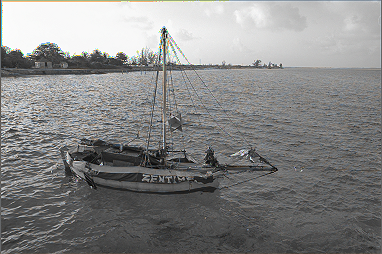}
    \caption{\label{fig:DebayerE0}original output}
  \end{subfigure}
  ~ 
  \begin{subfigure}{.44\linewidth}
    \includegraphics[width=\linewidth]{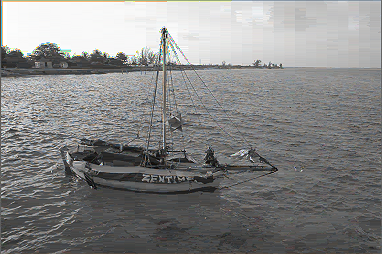}
    \caption{\label{fig:DebayerE3}output error \textless{} 3\%}
  \end{subfigure}
  \vspace{-0.3cm}
  \caption{\label{fig:DebayerOutput}Output quality of debayer.}
  \vspace{-0.2cm}
\end{figure}
\begin{figure}[!ht]
  \vspace{-0.2cm}
  \centering
  \begin{subfigure}{.44\linewidth}
    \includegraphics[width=\linewidth]{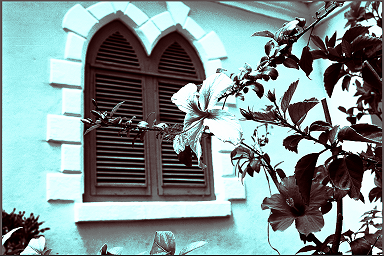}
    \caption{\label{fig:HisteqE0}original output}
  \end{subfigure}
  ~ 
  \begin{subfigure}{.44\linewidth}
    \includegraphics[width=\linewidth]{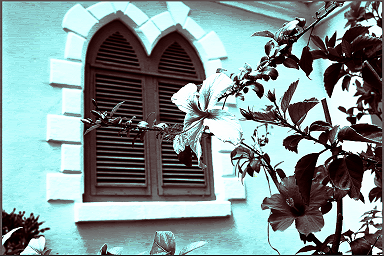}
    \caption{\label{fig:HisteqE3}output error \textless{} 3\%}
  \end{subfigure}
  \vspace{-0.3cm}
  \caption{\label{fig:HisteqOutput}Output quality of histeq.}
  \vspace{-0.5cm}
\end{figure}

Figures~\ref{fig:JpegOutput}-\ref{fig:HisteqOutput} show the output images 
subject to output error constraint of 3\% with the original images from precise 
computation~\cite{DBLP:conf/islped/RanjanRRR17}. Under 3\% quality loss,
the visual difference between relaxed output images and original images is slight.
The selection of error constraint may be different in
applications, which requires the knowledge on the accuracy requirements of applications. 
For the applications with high endurance for noises in
input data, such as feature extraction and machine learning, the error constraint
can be set relatively larger than the applications requiring high accuracy. 
Due to the difference
in image processing algorithms, the \emph{AF} for a
specific output error constraint varies in different workloads. A practical way to
determine the \emph{AF} is to search the suitable \emph{AF} using small canary inputs
and leverage the inferred \emph{AF} on full size inputs~\cite{DBLP:conf/pldi/LaurenzanoHSMMT16,
DBLP:conf/usenix/XuKKBMMB18}.

\subsection{Adaptability for Bitmap Format Variance}

\begin{figure}[!ht]
\centering
\includegraphics[width=0.4\textwidth]{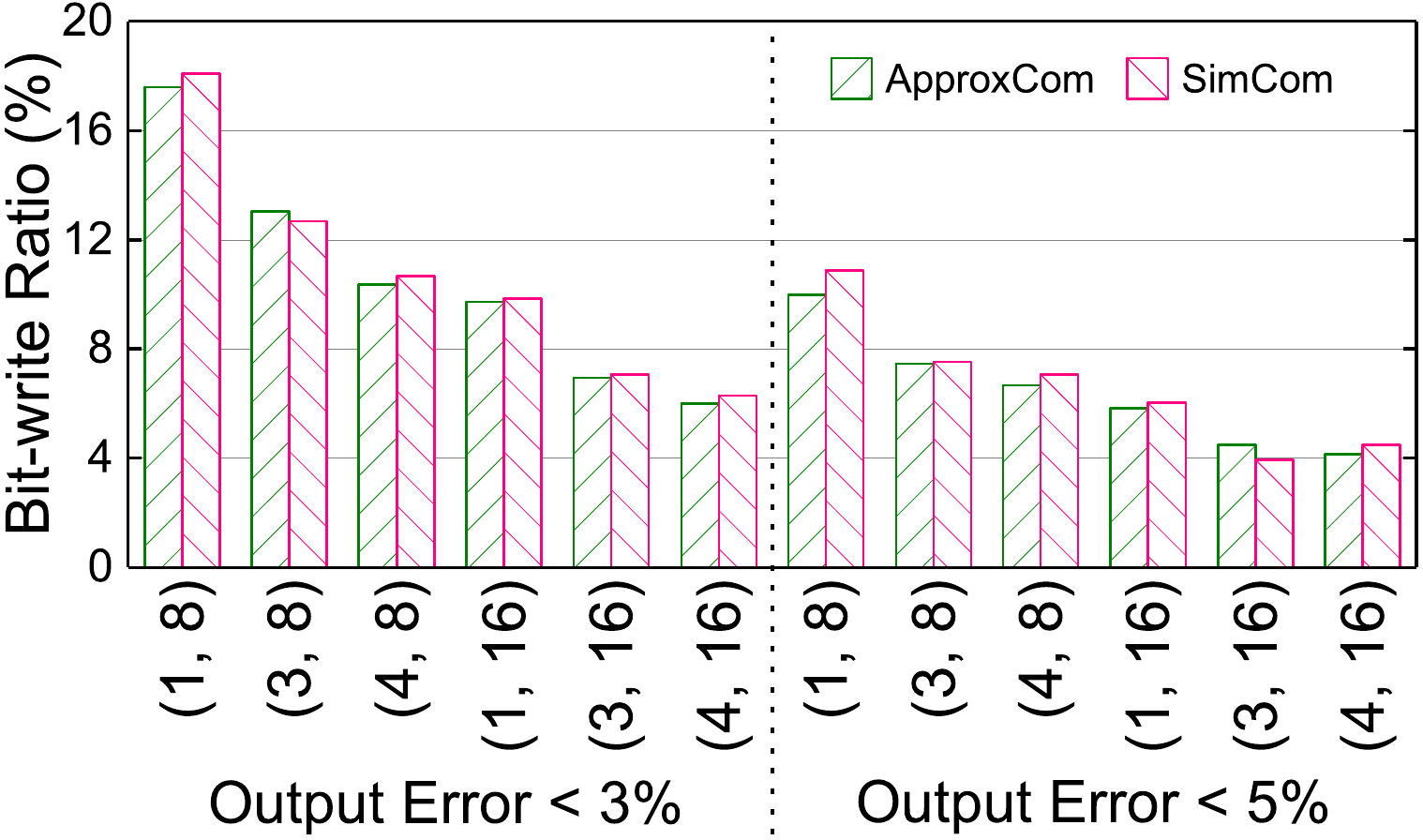}
\vspace{-0.2cm}
\caption{\label{fig:Adaptability}The bit-write ratio in jpeg with
bitmaps of different formats.
\emph{(A bitmap format of (m, n) indicates CC = m and BPC = n.)}}
\vspace{-0.4cm}
\end{figure}

In order to verify the adaptability of SimCom, we evaluate the jpeg workload
with input images of different formats. As shown in 
Figure~\ref{fig:Adaptability}, SimCom achieves comparable bit-write ratios
to those of ApproxCom (within 1\%). Without annotations on bitmap formats, SimCom
is able to infer the data types according to the mean 
difference (\secref{sec:AdaptiveCompressionScheme}) among data. 
The pixel-level similarity in data guarantees that the 
right compression mode (\secref{sec:AdaptiveCompression})
tends to obtain the minimal mean difference.

\begin{table}[!ht]
  \centering
  \caption{Statistics for the \emph{Mode Selector} in SimCom with 
    output error \textless{} 3\%.
  \emph{(The two integers in bitmap formats denote CC and BPC, respectively.)}}
  \vspace{-0.2cm}
  \label{tab:SelectorStat}
  \scalebox{0.8}{
    \begin{tabular}{|c|c|c|c|c|c|c|}
      \hline
      \multirow{2}{*}{Mode Ratio (\%)} & \multicolumn{6}{c|}{Bitmap Formats} \\
      \cline{2-7} & (1, 8) & (3, 8) & (4, 8) & (1, 16) & (2, 16) & (4, 16) \\ \hline
      1C1B & \textbf{82.4} & 47.2 & 0.6 & 0.2 & 0.2 & 0.2 \\ \hline
      3C1B & 0.2 & \textbf{34.1} & 0 & 0 & 0 & 0\\ \hline
      4C1B & 0.1 & 0.4 & \textbf{96.9} & 0 & 0 & 0\\ \hline
      1C2B & 15.3 & 7.4 & 0 & \textbf{98.5} & 58.3 & 1.0\\ \hline
      3C2B & 0 & 7.1 & 0 & 0.2 & \textbf{38.5} & 0\\ \hline
      4C2B & 0.1 & 0.1 & 2.2 & 0.6 & 1.6 & \textbf{97.9}\\ \hline
      Incompressible & 1.9 & 3.7 & 0.3 & 0.5 & 1.4 & 0.9\\ \hline
    \end{tabular}
  }
  \vspace{-0.2cm}
\end{table}

An interesting point is that SimCom obtains slightly lower bit-write
ratios when \emph{CC} is 3 
(e.g., bitmap formats of (3, 8) and (3, 16) when the
output error threshold is 3\% and 5\%, respectively)
otherwise marginally higher bit-write
ratios than ApproxCom. Since the \emph{Mode Selector} selects the 
compression mode with
minimal mean difference, it is possible to select a compression mode
with slightly smaller mean difference but much larger compressed data
size than the right compression mode. The conservative strategy used in SimCom
brings minor performance decrease, as shown in Figure~\ref{fig:Adaptability}. 
We record the selection of mode inside the \emph{Mode Selector} of SimCom.
Table~\ref{tab:SelectorStat} shows SimCom is able to obtain
the right compression mode in most cases (the numbers in boldface).
However, when \emph{CC} is 3,
the \emph{Mode Selector} possibly chooses the mode 
in which \emph{CC} is 1.
The reason is that when the values of 
three channels in a pixel are identical, e.g., pixels of white color 
and grayscale images stored in RGB formats, a compression mode with 
one channel can achieve the same mean difference with smaller 
compressed data size than the right compression mode. 
Therefore, SimCom achieves the adaptiveness in the mode selection
and low bit-write ratios for various bitmap formats.

\subsection{Discussion}

\textbf{The Principle of Approximate Compression Mode.}
There are six available compression modes in SimCom. 
Compression modes with other \emph{CC}s and \emph{BPC}s
can be added into SimCom like existing modes. For images 
with \emph{BPC} \textgreater{} 16, 
an alternative approach is to downscale the precision
to fit the images for predefined compression modes in SimCom. 
Due to the error-tolerance in images, slight
precision downscaling of images with large \emph{BPC} would not cause 
significant quality loss.

\textbf{Hardware Overhead of SimCom.}
The majority of hardware overhead of SimCom comes from the parallel execution
using six approximate compression logics, which can be optimized by 
reusing the logic.
In this case, six compression modes are executed one by one in the compression
logic, which trades the compression speed for the hardware efficiency.

\textbf{Architecture Support for SimCom.}
In the current testbed, SimCom requires architecture support 
(i.e., microarchitecture modifications and ISA extensions) 
like prior works~\cite{DBLP:conf/asplos/EsmaeilzadehSCB12,
DBLP:conf/micro/MiguelAMJ15,DBLP:conf/micro/MiguelAJJ16,
DBLP:conf/date/RanjanVPV0R17} to identify write accesses with 
approximable data, which are widely-used techniques in approximate 
storage systems~\cite{DBLP:conf/pldi/SampsonDFGCG11,
DBLP:conf/asplos/EsmaeilzadehSCB12,sampson2015accept,
DBLP:conf/micro/MiguelAMJ15,DBLP:conf/micro/MiguelAJJ16,
DBLP:conf/date/RanjanVPV0R17}. For image-based 
applications (e.g., machine learning and computer vision), memory 
performance and energy efficiency are important for entire system
performance. In addition, these applications are generally tolerant 
for minor errors. Moreover, power consumption is constrained in 
specific platforms (e.g., smartphones and embedded devices).
Therefore, it is meaningful to provide architecture support for 
approximate storage systems. 
With architecture support like Truffle~\cite{DBLP:conf/asplos/EsmaeilzadehSCB12}, SimCom 
only requires small hardware changes in the NVM module controller. 
In the meantime, it's possible to deliver accuracy requirements
via software interfaces 
without ISA extensions~\cite{DBLP:conf/islped/RanjanRRR17,
DBLP:conf/iccad/ZhaoXCZ17,DBLP:conf/asplos/LiuPMZ11}. 
Through the interfaces, approximable data are stored
in a separate memory region. Read or write 
accesses to the region can be identified by memory addresses, 
thus mitigating requirements for architecture support. 

\section{Related Work}
\label{sec:RelatedWork}



\textbf{Data Compression in NVMs.} 
Data compression schemes compress data to reduce the bits
written to NVMs.
FPC~\cite{DBLP:conf/nanoarch/DgienPHLM14} 
uses static data patterns to compress frequent
patterns into short prefix bits with remaining bits. However, the data patterns optimized
for the generalization in various applications don't match bitmaps, resulting
in poor compression performance for image-based applications. BDI~\cite{DBLP:conf/IEEEpact/PekhimenkoSMGKM12}
leverages the characteristics of narrow values in arrays to encode each word
using bases with small deltas. As shown in \secref{sec:Evaluation}, 
it's difficult for bitmaps to satisfy the constraints of FPC or BDI even 
with approximation. 
Different from FPC and BDI, SimCom leverages the pixel-level similarity 
in bitmaps and efficiently trades slight output quality for performance improvement. 



\textbf{Approximate Image Storage.}
To address the challenge of massive image collections, several approximation 
approaches are proposed to improve the efficiency of image storage. 
A biased MLC write scheme~\cite{DBLP:conf/asplos/GuoSCM16,DBLP:conf/asplos/JevdjicSCM17} 
is used to balance the drift and write errors in MLC PCM. Selective 
ECC is applied on images according to the importance of encoded bits~\cite{DBLP:conf/asplos/GuoSCM16}.
Progressively encoding scheme can improve the read performance of images~\cite{DBLP:conf/hotstorage/YanZWSC17}.
However, these schemes are established based on the significant entropy differences
in encoded image bits, which don't exist in bitmaps. Therefore, encoded image 
approximation is inefficient for the writes of bitmaps in NVMs.
Recent work~\cite{DBLP:conf/iccad/ZhaoXCZ17} proposes to selectively write pixels
in approximate window by writing soft bits in MLC STT-MRAM main memory. The
approximation is efficient when loading entire images from disks to MLC STT-MRAM.
However, it is limited to MLC STT-MRAM and needs searching for similar 
contents in other memory blocks, which leverages inter-block similarity 
and leads to additional hardware overheads and latency when writing 
data from cache to NVMs.

\textbf{Approximate Cache \& Main Memory.}
The accesses to memory can be served with predicted values according to
previous data patterns~\cite{DBLP:conf/micro/MiguelBJ14}. Doppelg{\"{a}}nger associates
similar data blocks with a single tag entry~\cite{DBLP:conf/micro/MiguelAMJ15}.
Bunker cache leverages the spatial-value similarity and maps similar data blocks 
to an identical cache entry~\cite{DBLP:conf/micro/MiguelAJJ16}.
The inter-block similarity used in the above caches is orthogonal to the 
pixel-level similarity of our work. Flikker reduces the refresh rates 
of DRAM portion containing error-tolerant 
data~\cite{DBLP:conf/asplos/LiuPMZ11}. Bidirectional precision scaling is proposed
to compress the data to be written to DRAM~\cite{DBLP:conf/islped/RanjanRRR17}. 
However, indiscriminately reducing the precision of all data can significantly 
decrease the image quality. Unlike them, SimCom exploits the pixel-level
similarity in bitmaps and efficiently reduce the write of similar words in NVMs
with minor quality loss.

\vspace{-0.4cm}
\section{Conclusion}
\label{sec:Conclusion}

Bit-write reduction in NVM is important for the performance of NVM-based
main memory. By exploiting the error-tolerance and similarity
in bitmaps, SimCom efficiently reduces the writes of similar words in 
write accesses to NVMs on-the-fly.
Due to the flexibility and efficiency of approximate compression, 
SimCom delivers higher performance than state-of-the-art compression schemes with
slight programmer annotations.

\vbadness=10000 
\hbadness=10000 

%
\bibliographystyle{ACM-Reference-Format}
\bibliography{../../../papers/NVM/nvm,../../../papers/Index/index,../../../papers/Approximation/approximation,../../../papers/Compression/compression,../../../papers/Security/security,../../../papers/ComputerVision/cv}


\begin{thebibliography}{53}


\ifx \showCODEN    \undefined \def \showCODEN     #1{\unskip}     \fi
\ifx \showDOI      \undefined \def \showDOI       #1{#1}\fi
\ifx \showISBNx    \undefined \def \showISBNx     #1{\unskip}     \fi
\ifx \showISBNxiii \undefined \def \showISBNxiii  #1{\unskip}     \fi
\ifx \showISSN     \undefined \def \showISSN      #1{\unskip}     \fi
\ifx \showLCCN     \undefined \def \showLCCN      #1{\unskip}     \fi
\ifx \shownote     \undefined \def \shownote      #1{#1}          \fi
\ifx \showarticletitle \undefined \def \showarticletitle #1{#1}   \fi
\ifx \showURL      \undefined \def \showURL       {\relax}        \fi
\providecommand\bibfield[2]{#2}
\providecommand\bibinfo[2]{#2}
\providecommand\natexlab[1]{#1}
\providecommand\showeprint[2][]{arXiv:#2}

\bibitem[\protect\citeauthoryear{Baek and Chilimbi}{Baek and Chilimbi}{2010}]%
        {DBLP:conf/pldi/BaekC10}
\bibfield{author}{\bibinfo{person}{Woongki Baek} {and}
  \bibinfo{person}{Trishul~M. Chilimbi}.} \bibinfo{year}{2010}\natexlab{}.
\newblock \showarticletitle{{Green: A Framework for Supporting Energy-Conscious
  Programming using Controlled Approximation}}. In
  \bibinfo{booktitle}{\emph{Proc. PLDI}}. \bibinfo{pages}{198--209}.
\newblock
\urldef\tempurl%
\url{https://doi.org/10.1145/1806596.1806620}
\showDOI{\tempurl}


\bibitem[\protect\citeauthoryear{Barker, Benson, Campbell, Ediger, Gioiosa,
  Hoisie, Kerbyson, Manzano, Marquez, Song, et~al\mbox{.}}{Barker
  et~al\mbox{.}}{2013}]%
        {barker2013perfect}
\bibfield{author}{\bibinfo{person}{Kevin Barker}, \bibinfo{person}{Thomas
  Benson}, \bibinfo{person}{Dan Campbell}, \bibinfo{person}{David Ediger},
  \bibinfo{person}{Roberto Gioiosa}, \bibinfo{person}{Adolfy Hoisie},
  \bibinfo{person}{Darren Kerbyson}, \bibinfo{person}{Joseph Manzano},
  \bibinfo{person}{Andres Marquez}, \bibinfo{person}{Leon Song},
  {et~al\mbox{.}}} \bibinfo{year}{2013}\natexlab{}.
\newblock \showarticletitle{{PERFECT (Power Efficiency Revolution For Embedded
  Computing Technologies) Benchmark Suite Manual}}.
\newblock \bibinfo{journal}{\emph{Pacific Northwest National Laboratory and
  Georgia Tech Research Institute}} (\bibinfo{year}{2013}).
\newblock


\bibitem[\protect\citeauthoryear{Binkert, Beckmann, Black, Reinhardt, Saidi,
  Basu, Hestness, Hower, Krishna, Sardashti, Sen, Sewell, Altaf, Vaish, Hill,
  and Wood}{Binkert et~al\mbox{.}}{2011}]%
        {DBLP:journals/sigarch/BinkertBBRSBHHKSSSSVHW11}
\bibfield{author}{\bibinfo{person}{Nathan~L. Binkert},
  \bibinfo{person}{Bradford~M. Beckmann}, \bibinfo{person}{Gabriel Black},
  \bibinfo{person}{Steven~K. Reinhardt}, \bibinfo{person}{Ali~G. Saidi},
  \bibinfo{person}{Arkaprava Basu}, \bibinfo{person}{Joel Hestness},
  \bibinfo{person}{Derek Hower}, \bibinfo{person}{Tushar Krishna},
  \bibinfo{person}{Somayeh Sardashti}, \bibinfo{person}{Rathijit Sen},
  \bibinfo{person}{Korey Sewell}, \bibinfo{person}{Muhammad Shoaib~Bin Altaf},
  \bibinfo{person}{Nilay Vaish}, \bibinfo{person}{Mark~D. Hill}, {and}
  \bibinfo{person}{David~A. Wood}.} \bibinfo{year}{2011}\natexlab{}.
\newblock \showarticletitle{{The gem5 Simulator}}.
\newblock \bibinfo{journal}{\emph{{SIGARCH} Computer Architecture News}}
  \bibinfo{volume}{39}, \bibinfo{number}{2} (\bibinfo{year}{2011}),
  \bibinfo{pages}{1--7}.
\newblock
\urldef\tempurl%
\url{https://doi.org/10.1145/2024716.2024718}
\showDOI{\tempurl}


\bibitem[\protect\citeauthoryear{Cho and Lee}{Cho and Lee}{2009}]%
        {DBLP:conf/micro/ChoL09}
\bibfield{author}{\bibinfo{person}{Sangyeun Cho} {and} \bibinfo{person}{Hyunjin
  Lee}.} \bibinfo{year}{2009}\natexlab{}.
\newblock \showarticletitle{{Flip-N-Write: A Simple Deterministic Technique to
  Improve PRAM Write Performance, Energy and Endurance}}. In
  \bibinfo{booktitle}{\emph{Proc. MICRO}}. \bibinfo{publisher}{{ACM}},
  \bibinfo{pages}{347--357}.
\newblock
\urldef\tempurl%
\url{https://doi.org/10.1145/1669112.1669157}
\showDOI{\tempurl}


\bibitem[\protect\citeauthoryear{Condit, Nightingale, Frost, Ipek, Lee, Burger,
  and Coetzee}{Condit et~al\mbox{.}}{2009}]%
        {DBLP:conf/sosp/ConditNFILBC09}
\bibfield{author}{\bibinfo{person}{Jeremy Condit}, \bibinfo{person}{Edmund~B.
  Nightingale}, \bibinfo{person}{Christopher Frost}, \bibinfo{person}{Engin
  Ipek}, \bibinfo{person}{Benjamin~C. Lee}, \bibinfo{person}{Doug Burger},
  {and} \bibinfo{person}{Derrick Coetzee}.} \bibinfo{year}{2009}\natexlab{}.
\newblock \showarticletitle{{Better I/O Through Byte-Addressable, Persistent
  Memory}}. In \bibinfo{booktitle}{\emph{Proc. SOSP}}.
  \bibinfo{publisher}{{ACM}}, \bibinfo{pages}{133--146}.
\newblock
\urldef\tempurl%
\url{https://doi.org/10.1145/1629575.1629589}
\showDOI{\tempurl}


\bibitem[\protect\citeauthoryear{David, Dragojevic, Guerraoui, and
  Zablotchi}{David et~al\mbox{.}}{2018}]%
        {DBLP:conf/usenix/DavidDGZ18}
\bibfield{author}{\bibinfo{person}{Tudor David}, \bibinfo{person}{Aleksandar
  Dragojevic}, \bibinfo{person}{Rachid Guerraoui}, {and} \bibinfo{person}{Igor
  Zablotchi}.} \bibinfo{year}{2018}\natexlab{}.
\newblock \showarticletitle{{Log-Free Concurrent Data Structures}}. In
  \bibinfo{booktitle}{\emph{Proc. ATC}}. \bibinfo{publisher}{{USENIX}
  Association}, \bibinfo{pages}{373--386}.
\newblock
\urldef\tempurl%
\url{https://www.usenix.org/conference/atc18/presentation/david}
\showURL{%
\tempurl}


\bibitem[\protect\citeauthoryear{Dgien, Palangappa, Hunter, Li, and
  Mohanram}{Dgien et~al\mbox{.}}{2014}]%
        {DBLP:conf/nanoarch/DgienPHLM14}
\bibfield{author}{\bibinfo{person}{David~B. Dgien},
  \bibinfo{person}{Poovaiah~M. Palangappa}, \bibinfo{person}{Nathan~Altay
  Hunter}, \bibinfo{person}{Jiayin Li}, {and} \bibinfo{person}{Kartik
  Mohanram}.} \bibinfo{year}{2014}\natexlab{}.
\newblock \showarticletitle{{Compression Architecture for Bit-write Reduction
  in Non-volatile Memory Technologies}}. In \bibinfo{booktitle}{\emph{Proc.
  NANOARCH}}. \bibinfo{publisher}{ACM}, \bibinfo{pages}{51--56}.
\newblock
\urldef\tempurl%
\url{https://doi.org/10.1109/NANOARCH.2014.6880482}
\showDOI{\tempurl}


\bibitem[\protect\citeauthoryear{Dufaux, Sullivan, and Ebrahimi}{Dufaux
  et~al\mbox{.}}{2009}]%
        {dufaux2009jpeg}
\bibfield{author}{\bibinfo{person}{Fr{\'e}d{\'e}ric Dufaux},
  \bibinfo{person}{Gary~J Sullivan}, {and} \bibinfo{person}{Touradj Ebrahimi}.}
  \bibinfo{year}{2009}\natexlab{}.
\newblock \showarticletitle{{The JPEG XR Image Coding Standard [Standards in a
  Nutshell]}}.
\newblock \bibinfo{journal}{\emph{IEEE Signal Processing Magazine}}
  \bibinfo{volume}{26}, \bibinfo{number}{6} (\bibinfo{year}{2009}).
\newblock


\bibitem[\protect\citeauthoryear{Duff}{Duff}{2017}]%
        {DBLP:journals/tog/Duff17}
\bibfield{author}{\bibinfo{person}{Tom Duff}.} \bibinfo{year}{2017}\natexlab{}.
\newblock \showarticletitle{Deep Compositing Using Lie Algebras}.
\newblock \bibinfo{journal}{\emph{{ACM} Trans. Graph.}} \bibinfo{volume}{36},
  \bibinfo{number}{3} (\bibinfo{year}{2017}), \bibinfo{pages}{26:1--26:12}.
\newblock
\urldef\tempurl%
\url{https://doi.org/10.1145/3023386}
\showDOI{\tempurl}


\bibitem[\protect\citeauthoryear{Esmaeilzadeh, Sampson, Ceze, and
  Burger}{Esmaeilzadeh et~al\mbox{.}}{2012}]%
        {DBLP:conf/asplos/EsmaeilzadehSCB12}
\bibfield{author}{\bibinfo{person}{Hadi Esmaeilzadeh}, \bibinfo{person}{Adrian
  Sampson}, \bibinfo{person}{Luis Ceze}, {and} \bibinfo{person}{Doug Burger}.}
  \bibinfo{year}{2012}\natexlab{}.
\newblock \showarticletitle{{Architecture Support for Disciplined Approximate
  Programming}}. In \bibinfo{booktitle}{\emph{Proc. ASPLOS}}.
  \bibinfo{pages}{301--312}.
\newblock
\urldef\tempurl%
\url{https://doi.org/10.1145/2150976.2151008}
\showDOI{\tempurl}


\bibitem[\protect\citeauthoryear{Franzen}{Franzen}{[n. d.]}]%
        {kodakimage}
\bibfield{author}{\bibinfo{person}{Rich Franzen}.} \bibinfo{year}{[n.
  d.]}\natexlab{}.
\newblock \bibinfo{title}{{Kodak Lossless True Color Image Suite}}.
\newblock
\newblock
\urldef\tempurl%
\url{http://r0k.us/graphics/kodak/}
\showURL{%
Retrieved April 10, 2019 from \tempurl}


\bibitem[\protect\citeauthoryear{Guo, Strauss, Ceze, and Malvar}{Guo
  et~al\mbox{.}}{2016}]%
        {DBLP:conf/asplos/GuoSCM16}
\bibfield{author}{\bibinfo{person}{Qing Guo}, \bibinfo{person}{Karin Strauss},
  \bibinfo{person}{Luis Ceze}, {and} \bibinfo{person}{Henrique~S. Malvar}.}
  \bibinfo{year}{2016}\natexlab{}.
\newblock \showarticletitle{{High-Density Image Storage Using Approximate
  Memory Cells}}. In \bibinfo{booktitle}{\emph{Proc. ASPLOS}}.
  \bibinfo{pages}{413--426}.
\newblock
\urldef\tempurl%
\url{https://doi.org/10.1145/2872362.2872413}
\showDOI{\tempurl}


\bibitem[\protect\citeauthoryear{Guo, Hua, and Zuo}{Guo et~al\mbox{.}}{2018}]%
        {DBLP:conf/date/GuoHZ18}
\bibfield{author}{\bibinfo{person}{Yuncheng Guo}, \bibinfo{person}{Yu Hua},
  {and} \bibinfo{person}{Pengfei Zuo}.} \bibinfo{year}{2018}\natexlab{}.
\newblock \showarticletitle{{DFPC: A Dynamic Frequent Pattern Compression
  Scheme in NVM-based Main Memory}}. In \bibinfo{booktitle}{\emph{Proc. DATE}}.
  \bibinfo{publisher}{{IEEE}}, \bibinfo{pages}{1622--1627}.
\newblock
\urldef\tempurl%
\url{https://doi.org/10.23919/DATE.2018.8342274}
\showDOI{\tempurl}


\bibitem[\protect\citeauthoryear{Hong, Nair, Abali, Buyuktosunoglu, Kim, and
  Healy}{Hong et~al\mbox{.}}{2018}]%
        {DBLP:conf/micro/HongNABKH18}
\bibfield{author}{\bibinfo{person}{Seokin Hong},
  \bibinfo{person}{Prashant~Jayaprakash Nair}, \bibinfo{person}{B{\"{u}}lent
  Abali}, \bibinfo{person}{Alper Buyuktosunoglu}, \bibinfo{person}{Kyu{-}Hyoun
  Kim}, {and} \bibinfo{person}{Michael~B. Healy}.}
  \bibinfo{year}{2018}\natexlab{}.
\newblock \showarticletitle{Attach{\'{e}}: Towards Ideal Memory Compression by
  Mitigating Metadata Bandwidth Overheads}. In \bibinfo{booktitle}{\emph{Proc.
  MICRO}}. \bibinfo{publisher}{{IEEE}}, \bibinfo{pages}{326--338}.
\newblock
\urldef\tempurl%
\url{https://doi.org/10.1109/MICRO.2018.00034}
\showDOI{\tempurl}


\bibitem[\protect\citeauthoryear{Hwang, Kim, Won, and Nam}{Hwang
  et~al\mbox{.}}{2018}]%
        {DBLP:conf/fast/HwangKWN18}
\bibfield{author}{\bibinfo{person}{Deukyeon Hwang}, \bibinfo{person}{Wook{-}Hee
  Kim}, \bibinfo{person}{Youjip Won}, {and} \bibinfo{person}{Beomseok Nam}.}
  \bibinfo{year}{2018}\natexlab{}.
\newblock \showarticletitle{{Endurable Transient Inconsistency in
  Byte-Addressable Persistent B+-Tree}}. In \bibinfo{booktitle}{\emph{Proc.
  FAST}}. \bibinfo{publisher}{{USENIX} Association}, \bibinfo{pages}{187--200}.
\newblock
\urldef\tempurl%
\url{https://www.usenix.org/conference/fast18/presentation/hwang}
\showURL{%
\tempurl}


\bibitem[\protect\citeauthoryear{Jacobvitz, Calderbank, and Sorin}{Jacobvitz
  et~al\mbox{.}}{2013}]%
        {DBLP:conf/hpca/JacobvitzCS13}
\bibfield{author}{\bibinfo{person}{Adam~N. Jacobvitz},
  \bibinfo{person}{A.~Robert Calderbank}, {and} \bibinfo{person}{Daniel~J.
  Sorin}.} \bibinfo{year}{2013}\natexlab{}.
\newblock \showarticletitle{{Coset Coding to Extend the Lifetime of Memory}}.
  In \bibinfo{booktitle}{\emph{Proc. HPCA}}. \bibinfo{publisher}{{IEEE}
  Computer Society}, \bibinfo{pages}{222--233}.
\newblock
\urldef\tempurl%
\url{https://doi.org/10.1109/HPCA.2013.6522321}
\showDOI{\tempurl}


\bibitem[\protect\citeauthoryear{Jevdjic, Strauss, Ceze, and Malvar}{Jevdjic
  et~al\mbox{.}}{2017}]%
        {DBLP:conf/asplos/JevdjicSCM17}
\bibfield{author}{\bibinfo{person}{Djordje Jevdjic}, \bibinfo{person}{Karin
  Strauss}, \bibinfo{person}{Luis Ceze}, {and} \bibinfo{person}{Henrique~S.
  Malvar}.} \bibinfo{year}{2017}\natexlab{}.
\newblock \showarticletitle{{Approximate Storage of Compressed and Encrypted
  Videos}}. In \bibinfo{booktitle}{\emph{Proc. ASPLOS}}.
  \bibinfo{pages}{361--373}.
\newblock
\urldef\tempurl%
\url{https://doi.org/10.1145/3037697.3037718}
\showDOI{\tempurl}


\bibitem[\protect\citeauthoryear{Judd}{Judd}{1975}]%
        {judd1975color}
\bibfield{author}{\bibinfo{person}{Deane~B Judd}.}
  \bibinfo{year}{1975}\natexlab{}.
\newblock \bibinfo{booktitle}{\emph{{Color in Business, Science and
  Industry}}}.
\newblock \bibinfo{publisher}{Wiley-Interscience}.
\newblock


\bibitem[\protect\citeauthoryear{Laurenzano, Hill, Samadi, Mahlke, Mars, and
  Tang}{Laurenzano et~al\mbox{.}}{2016}]%
        {DBLP:conf/pldi/LaurenzanoHSMMT16}
\bibfield{author}{\bibinfo{person}{Michael~A. Laurenzano},
  \bibinfo{person}{Parker Hill}, \bibinfo{person}{Mehrzad Samadi},
  \bibinfo{person}{Scott~A. Mahlke}, \bibinfo{person}{Jason Mars}, {and}
  \bibinfo{person}{Lingjia Tang}.} \bibinfo{year}{2016}\natexlab{}.
\newblock \showarticletitle{{Input Responsiveness: Using Canary Inputs to
  Dynamically Steer Approximation}}. In \bibinfo{booktitle}{\emph{Proc. PLDI}}.
  \bibinfo{pages}{161--176}.
\newblock
\urldef\tempurl%
\url{https://doi.org/10.1145/2908080.2908087}
\showDOI{\tempurl}


\bibitem[\protect\citeauthoryear{Lee, Ipek, Mutlu, and Burger}{Lee
  et~al\mbox{.}}{2009}]%
        {DBLP:conf/isca/LeeIMB09}
\bibfield{author}{\bibinfo{person}{Benjamin~C. Lee}, \bibinfo{person}{Engin
  Ipek}, \bibinfo{person}{Onur Mutlu}, {and} \bibinfo{person}{Doug Burger}.}
  \bibinfo{year}{2009}\natexlab{}.
\newblock \showarticletitle{{Architecting Phase Change Memory as a Scalable
  DRAM Alternative}}. In \bibinfo{booktitle}{\emph{Proc. ISCA}}.
  \bibinfo{publisher}{{ACM}}, \bibinfo{pages}{2--13}.
\newblock
\urldef\tempurl%
\url{https://doi.org/10.1145/1555754.1555758}
\showDOI{\tempurl}


\bibitem[\protect\citeauthoryear{Leong}{Leong}{2006}]%
        {leong2006number}
\bibfield{author}{\bibinfo{person}{J Leong}.} \bibinfo{year}{2006}\natexlab{}.
\newblock \showarticletitle{{Number of colors distinguishable by the human
  eye}}.
\newblock \bibinfo{journal}{\emph{Hypertextbook,(ed.). Wyszecki, Gunter. Color.
  Chicago: World Book Inc}}  \bibinfo{volume}{824} (\bibinfo{year}{2006}).
\newblock


\bibitem[\protect\citeauthoryear{Li, Zhou, and Li}{Li et~al\mbox{.}}{2013}]%
        {DBLP:conf/hpca/LiZL13a}
\bibfield{author}{\bibinfo{person}{Zhongqi Li}, \bibinfo{person}{Ruijin Zhou},
  {and} \bibinfo{person}{Tao Li}.} \bibinfo{year}{2013}\natexlab{}.
\newblock \showarticletitle{{Exploring High-Performance and Energy Proportional
  Interface for Phase Change Memory Systems}}. In
  \bibinfo{booktitle}{\emph{Proc. HPCA}}. \bibinfo{publisher}{{IEEE} Computer
  Society}, \bibinfo{pages}{210--221}.
\newblock
\urldef\tempurl%
\url{https://doi.org/10.1109/HPCA.2013.6522320}
\showDOI{\tempurl}


\bibitem[\protect\citeauthoryear{Liu, Pattabiraman, Moscibroda, and Zorn}{Liu
  et~al\mbox{.}}{2011}]%
        {DBLP:conf/asplos/LiuPMZ11}
\bibfield{author}{\bibinfo{person}{Song Liu}, \bibinfo{person}{Karthik
  Pattabiraman}, \bibinfo{person}{Thomas Moscibroda}, {and}
  \bibinfo{person}{Benjamin~G. Zorn}.} \bibinfo{year}{2011}\natexlab{}.
\newblock \showarticletitle{{Flikker: Saving DRAM Refresh-power through
  Critical Data Partitioning}}. In \bibinfo{booktitle}{\emph{Proc. ASPLOS}}.
  \bibinfo{pages}{213--224}.
\newblock
\urldef\tempurl%
\url{https://doi.org/10.1145/1950365.1950391}
\showDOI{\tempurl}


\bibitem[\protect\citeauthoryear{Lowe}{Lowe}{2004}]%
        {DBLP:journals/ijcv/Lowe04}
\bibfield{author}{\bibinfo{person}{David~G. Lowe}.}
  \bibinfo{year}{2004}\natexlab{}.
\newblock \showarticletitle{{Distinctive Image Features from Scale-Invariant
  Keypoints}}.
\newblock \bibinfo{journal}{\emph{International Journal of Computer Vision}}
  \bibinfo{volume}{60}, \bibinfo{number}{2} (\bibinfo{year}{2004}),
  \bibinfo{pages}{91--110}.
\newblock
\urldef\tempurl%
\url{https://doi.org/10.1023/B:VISI.0000029664.99615.94}
\showDOI{\tempurl}


\bibitem[\protect\citeauthoryear{Miguel, Albericio, Jerger, and Jaleel}{Miguel
  et~al\mbox{.}}{2016}]%
        {DBLP:conf/micro/MiguelAJJ16}
\bibfield{author}{\bibinfo{person}{Joshua~San Miguel}, \bibinfo{person}{Jorge
  Albericio}, \bibinfo{person}{Natalie D.~Enright Jerger}, {and}
  \bibinfo{person}{Aamer Jaleel}.} \bibinfo{year}{2016}\natexlab{}.
\newblock \showarticletitle{{The Bunker Cache for Spatio-Value Approximation}}.
  In \bibinfo{booktitle}{\emph{Proc. MICRO}}. \bibinfo{pages}{43:1--43:12}.
\newblock
\urldef\tempurl%
\url{https://doi.org/10.1109/MICRO.2016.7783746}
\showDOI{\tempurl}


\bibitem[\protect\citeauthoryear{Miguel, Albericio, Moshovos, and
  Jerger}{Miguel et~al\mbox{.}}{2015}]%
        {DBLP:conf/micro/MiguelAMJ15}
\bibfield{author}{\bibinfo{person}{Joshua~San Miguel}, \bibinfo{person}{Jorge
  Albericio}, \bibinfo{person}{Andreas Moshovos}, {and}
  \bibinfo{person}{Natalie D.~Enright Jerger}.}
  \bibinfo{year}{2015}\natexlab{}.
\newblock \showarticletitle{{Doppelg{\"{a}}nger: A Cache for Approximate
  Computing}}. In \bibinfo{booktitle}{\emph{Proc. MICRO}}.
  \bibinfo{pages}{50--61}.
\newblock
\urldef\tempurl%
\url{https://doi.org/10.1145/2830772.2830790}
\showDOI{\tempurl}


\bibitem[\protect\citeauthoryear{Miguel, Badr, and Jerger}{Miguel
  et~al\mbox{.}}{2014}]%
        {DBLP:conf/micro/MiguelBJ14}
\bibfield{author}{\bibinfo{person}{Joshua~San Miguel}, \bibinfo{person}{Mario
  Badr}, {and} \bibinfo{person}{Natalie D.~Enright Jerger}.}
  \bibinfo{year}{2014}\natexlab{}.
\newblock \showarticletitle{{Load Value Approximation}}. In
  \bibinfo{booktitle}{\emph{Proc. MICRO}}. \bibinfo{pages}{127--139}.
\newblock
\urldef\tempurl%
\url{https://doi.org/10.1109/MICRO.2014.22}
\showDOI{\tempurl}


\bibitem[\protect\citeauthoryear{Palangappa and Mohanram}{Palangappa and
  Mohanram}{2016}]%
        {DBLP:conf/hpca/PalangappaM16}
\bibfield{author}{\bibinfo{person}{Poovaiah~M. Palangappa} {and}
  \bibinfo{person}{Kartik Mohanram}.} \bibinfo{year}{2016}\natexlab{}.
\newblock \showarticletitle{{CompEx: Compression-Expansion Coding for Energy,
  Latency, and Lifetime Improvements in MLC/TLC NVM}}. In
  \bibinfo{booktitle}{\emph{Proc. HPCA}}. \bibinfo{publisher}{{IEEE} Computer
  Society}, \bibinfo{pages}{90--101}.
\newblock
\urldef\tempurl%
\url{https://doi.org/10.1109/HPCA.2016.7446056}
\showDOI{\tempurl}


\bibitem[\protect\citeauthoryear{Palangappa and Mohanram}{Palangappa and
  Mohanram}{2018}]%
        {DBLP:conf/dac/PalangappaM18}
\bibfield{author}{\bibinfo{person}{Poovaiah~M. Palangappa} {and}
  \bibinfo{person}{Kartik Mohanram}.} \bibinfo{year}{2018}\natexlab{}.
\newblock \showarticletitle{{CASTLE: Compression Architecture for Secure Low
  Latency, Low Energy, High Endurance NVMs}}. In
  \bibinfo{booktitle}{\emph{Proc. DAC}}. \bibinfo{publisher}{{ACM}},
  \bibinfo{pages}{87:1--87:6}.
\newblock
\urldef\tempurl%
\url{https://doi.org/10.1145/3195970.3196007}
\showDOI{\tempurl}


\bibitem[\protect\citeauthoryear{Pekhimenko, Seshadri, Mutlu, Gibbons, Kozuch,
  and Mowry}{Pekhimenko et~al\mbox{.}}{2012}]%
        {DBLP:conf/IEEEpact/PekhimenkoSMGKM12}
\bibfield{author}{\bibinfo{person}{Gennady Pekhimenko}, \bibinfo{person}{Vivek
  Seshadri}, \bibinfo{person}{Onur Mutlu}, \bibinfo{person}{Phillip~B.
  Gibbons}, \bibinfo{person}{Michael~A. Kozuch}, {and} \bibinfo{person}{Todd~C.
  Mowry}.} \bibinfo{year}{2012}\natexlab{}.
\newblock \showarticletitle{{Base-Delta-Immediate Compression: Practical Data
  Compression for On-Chip Caches}}. In \bibinfo{booktitle}{\emph{Proc. PACT}}.
  \bibinfo{pages}{377--388}.
\newblock
\urldef\tempurl%
\url{https://doi.org/10.1145/2370816.2370870}
\showDOI{\tempurl}


\bibitem[\protect\citeauthoryear{Poremba, Zhang, and Xie}{Poremba
  et~al\mbox{.}}{2015}]%
        {DBLP:journals/cal/PorembaZ015}
\bibfield{author}{\bibinfo{person}{Matthew Poremba}, \bibinfo{person}{Tao
  Zhang}, {and} \bibinfo{person}{Yuan Xie}.} \bibinfo{year}{2015}\natexlab{}.
\newblock \showarticletitle{{NVMain 2.0: A User-Friendly Memory Simulator to
  Model (Non-)Volatile Memory Systems}}.
\newblock \bibinfo{journal}{\emph{CAL}} \bibinfo{volume}{14},
  \bibinfo{number}{2} (\bibinfo{year}{2015}), \bibinfo{pages}{140--143}.
\newblock
\urldef\tempurl%
\url{https://doi.org/10.1109/LCA.2015.2402435}
\showDOI{\tempurl}


\bibitem[\protect\citeauthoryear{Porter and Duff}{Porter and Duff}{1984}]%
        {DBLP:conf/siggraph/PorterD84}
\bibfield{author}{\bibinfo{person}{Thomas~K. Porter} {and} \bibinfo{person}{Tom
  Duff}.} \bibinfo{year}{1984}\natexlab{}.
\newblock \showarticletitle{{Compositing Digital Images}}. In
  \bibinfo{booktitle}{\emph{Proc. SIGGRAPH}}. \bibinfo{publisher}{{ACM}},
  \bibinfo{pages}{253--259}.
\newblock
\urldef\tempurl%
\url{https://doi.org/10.1145/800031.808606}
\showDOI{\tempurl}


\bibitem[\protect\citeauthoryear{Ranjan, Raha, Raghunathan, and
  Raghunathan}{Ranjan et~al\mbox{.}}{2017a}]%
        {DBLP:conf/islped/RanjanRRR17}
\bibfield{author}{\bibinfo{person}{Ashish Ranjan}, \bibinfo{person}{Arnab
  Raha}, \bibinfo{person}{Vijay Raghunathan}, {and} \bibinfo{person}{Anand
  Raghunathan}.} \bibinfo{year}{2017}\natexlab{a}.
\newblock \showarticletitle{{Approximate Memory Compression for
  Energy-efficiency}}. In \bibinfo{booktitle}{\emph{Proc. ISLPED}}.
  \bibinfo{pages}{1--6}.
\newblock
\urldef\tempurl%
\url{https://doi.org/10.1109/ISLPED.2017.8009173}
\showDOI{\tempurl}


\bibitem[\protect\citeauthoryear{Ranjan, Venkataramani, Pajouhi, Venkatesan,
  Roy, and Raghunathan}{Ranjan et~al\mbox{.}}{2017b}]%
        {DBLP:conf/date/RanjanVPV0R17}
\bibfield{author}{\bibinfo{person}{Ashish Ranjan}, \bibinfo{person}{Swagath
  Venkataramani}, \bibinfo{person}{Zoha Pajouhi}, \bibinfo{person}{Rangharajan
  Venkatesan}, \bibinfo{person}{Kaushik Roy}, {and} \bibinfo{person}{Anand
  Raghunathan}.} \bibinfo{year}{2017}\natexlab{b}.
\newblock \showarticletitle{{STAxCache: An Approximate, Energy Efficient
  STT-MRAM Cache}}. In \bibinfo{booktitle}{\emph{Proc. DATE}}.
  \bibinfo{pages}{356--361}.
\newblock
\urldef\tempurl%
\url{https://doi.org/10.23919/DATE.2017.7927016}
\showDOI{\tempurl}


\bibitem[\protect\citeauthoryear{Rublee, Rabaud, Konolige, and Bradski}{Rublee
  et~al\mbox{.}}{2011}]%
        {DBLP:conf/iccv/RubleeRKB11}
\bibfield{author}{\bibinfo{person}{Ethan Rublee}, \bibinfo{person}{Vincent
  Rabaud}, \bibinfo{person}{Kurt Konolige}, {and} \bibinfo{person}{Gary~R.
  Bradski}.} \bibinfo{year}{2011}\natexlab{}.
\newblock \showarticletitle{{ORB: an efficient alternative to SIFT or SURF}}.
  In \bibinfo{booktitle}{\emph{Proc. ICCV}}. \bibinfo{publisher}{{IEEE}
  Computer Society}, \bibinfo{pages}{2564--2571}.
\newblock
\urldef\tempurl%
\url{https://doi.org/10.1109/ICCV.2011.6126544}
\showDOI{\tempurl}


\bibitem[\protect\citeauthoryear{Samadi, Jamshidi, Lee, and Mahlke}{Samadi
  et~al\mbox{.}}{2014}]%
        {DBLP:conf/asplos/SamadiJLM14}
\bibfield{author}{\bibinfo{person}{Mehrzad Samadi},
  \bibinfo{person}{Davoud~Anoushe Jamshidi}, \bibinfo{person}{Janghaeng Lee},
  {and} \bibinfo{person}{Scott~A. Mahlke}.} \bibinfo{year}{2014}\natexlab{}.
\newblock \showarticletitle{{Paraprox: Pattern-Based Approximation for Data
  Parallel Applications}}. In \bibinfo{booktitle}{\emph{Proc. ASPLOS}}.
  \bibinfo{pages}{35--50}.
\newblock
\urldef\tempurl%
\url{https://doi.org/10.1145/2541940.2541948}
\showDOI{\tempurl}


\bibitem[\protect\citeauthoryear{Sampson, Baixo, Ransford, Moreau, Yip, Ceze,
  and Oskin}{Sampson et~al\mbox{.}}{2015}]%
        {sampson2015accept}
\bibfield{author}{\bibinfo{person}{Adrian Sampson}, \bibinfo{person}{Andr{\'e}
  Baixo}, \bibinfo{person}{Benjamin Ransford}, \bibinfo{person}{Thierry
  Moreau}, \bibinfo{person}{Joshua Yip}, \bibinfo{person}{Luis Ceze}, {and}
  \bibinfo{person}{Mark Oskin}.} \bibinfo{year}{2015}\natexlab{}.
\newblock \showarticletitle{{ACCEPT: A Programmer-Guided Compiler Framework for
  Practical Approximate Computing}}.
\newblock \bibinfo{journal}{\emph{University of Washington Technical Report
  UW-CSE-15-01}}  \bibinfo{volume}{1} (\bibinfo{year}{2015}).
\newblock


\bibitem[\protect\citeauthoryear{Sampson, Dietl, Fortuna, Gnanapragasam, Ceze,
  and Grossman}{Sampson et~al\mbox{.}}{2011}]%
        {DBLP:conf/pldi/SampsonDFGCG11}
\bibfield{author}{\bibinfo{person}{Adrian Sampson}, \bibinfo{person}{Werner
  Dietl}, \bibinfo{person}{Emily Fortuna}, \bibinfo{person}{Danushen
  Gnanapragasam}, \bibinfo{person}{Luis Ceze}, {and} \bibinfo{person}{Dan
  Grossman}.} \bibinfo{year}{2011}\natexlab{}.
\newblock \showarticletitle{{EnerJ: Approximate Data Types for Safe and General
  Low-Power Computation}}. In \bibinfo{booktitle}{\emph{Proc. PLDI}}.
  \bibinfo{pages}{164--174}.
\newblock
\urldef\tempurl%
\url{https://doi.org/10.1145/1993498.1993518}
\showDOI{\tempurl}


\bibitem[\protect\citeauthoryear{Sampson, Nelson, Strauss, and Ceze}{Sampson
  et~al\mbox{.}}{2013}]%
        {DBLP:conf/micro/SampsonNSC13}
\bibfield{author}{\bibinfo{person}{Adrian Sampson}, \bibinfo{person}{Jacob
  Nelson}, \bibinfo{person}{Karin Strauss}, {and} \bibinfo{person}{Luis Ceze}.}
  \bibinfo{year}{2013}\natexlab{}.
\newblock \showarticletitle{{Approximate Storage in Solid-State Memories}}. In
  \bibinfo{booktitle}{\emph{Proc. MICRO}}. \bibinfo{pages}{25--36}.
\newblock
\urldef\tempurl%
\url{https://doi.org/10.1145/2540708.2540712}
\showDOI{\tempurl}


\bibitem[\protect\citeauthoryear{Shin, Tirukkovalluri, Tuck, and Solihin}{Shin
  et~al\mbox{.}}{2017}]%
        {DBLP:conf/micro/ShinTTS17}
\bibfield{author}{\bibinfo{person}{Seunghee Shin},
  \bibinfo{person}{Satish~Kumar Tirukkovalluri}, \bibinfo{person}{James Tuck},
  {and} \bibinfo{person}{Yan Solihin}.} \bibinfo{year}{2017}\natexlab{}.
\newblock \showarticletitle{{Proteus: A Flexible and Fast Software Supported
  Hardware Logging approachs for NVM}}. In \bibinfo{booktitle}{\emph{Proc.
  MICRO}}. \bibinfo{publisher}{{ACM}}, \bibinfo{pages}{178--190}.
\newblock
\urldef\tempurl%
\url{https://doi.org/10.1145/3123939.3124539}
\showDOI{\tempurl}


\bibitem[\protect\citeauthoryear{Sui, Lenharth, Fussell, and Pingali}{Sui
  et~al\mbox{.}}{2016}]%
        {DBLP:conf/asplos/SuiLFP16}
\bibfield{author}{\bibinfo{person}{Xin Sui}, \bibinfo{person}{Andrew Lenharth},
  \bibinfo{person}{Donald~S. Fussell}, {and} \bibinfo{person}{Keshav Pingali}.}
  \bibinfo{year}{2016}\natexlab{}.
\newblock \showarticletitle{{Proactive Control of Approximate Programs}}. In
  \bibinfo{booktitle}{\emph{Proc. ASPLOS}}. \bibinfo{publisher}{{ACM}},
  \bibinfo{pages}{607--621}.
\newblock
\urldef\tempurl%
\url{https://doi.org/10.1145/2872362.2872402}
\showDOI{\tempurl}


\bibitem[\protect\citeauthoryear{Wallace}{Wallace}{1991}]%
        {DBLP:journals/cacm/Wallace91}
\bibfield{author}{\bibinfo{person}{Gregory~K. Wallace}.}
  \bibinfo{year}{1991}\natexlab{}.
\newblock \showarticletitle{{The JPEG Still Picture Compression Standard}}.
\newblock \bibinfo{journal}{\emph{Commun. {ACM}}} \bibinfo{volume}{34},
  \bibinfo{number}{4} (\bibinfo{year}{1991}), \bibinfo{pages}{30--44}.
\newblock
\urldef\tempurl%
\url{https://doi.org/10.1145/103085.103089}
\showDOI{\tempurl}


\bibitem[\protect\citeauthoryear{Xu, Feng, Hua, Tong, Liu, and Li}{Xu
  et~al\mbox{.}}{2018a}]%
        {DBLP:conf/date/XuFHTLL18}
\bibfield{author}{\bibinfo{person}{Jie Xu}, \bibinfo{person}{Dan Feng},
  \bibinfo{person}{Yu Hua}, \bibinfo{person}{Wei Tong},
  \bibinfo{person}{Jingning Liu}, {and} \bibinfo{person}{Chunyan Li}.}
  \bibinfo{year}{2018}\natexlab{a}.
\newblock \showarticletitle{{Extending the Lifetime of NVMs with Compression}}.
  In \bibinfo{booktitle}{\emph{Proc. DATE}}. \bibinfo{publisher}{{IEEE}},
  \bibinfo{pages}{1604--1609}.
\newblock
\urldef\tempurl%
\url{https://doi.org/10.23919/DATE.2018.8342271}
\showDOI{\tempurl}


\bibitem[\protect\citeauthoryear{Xu, Zhang, Memaripour, Gangadharaiah, Borase,
  Silva, Swanson, and Rudoff}{Xu et~al\mbox{.}}{2017}]%
        {DBLP:conf/sosp/XuZMGBSSR17}
\bibfield{author}{\bibinfo{person}{Jian Xu}, \bibinfo{person}{Lu Zhang},
  \bibinfo{person}{Amirsaman Memaripour}, \bibinfo{person}{Akshatha
  Gangadharaiah}, \bibinfo{person}{Amit Borase}, \bibinfo{person}{Tamires
  Brito~Da Silva}, \bibinfo{person}{Steven Swanson}, {and}
  \bibinfo{person}{Andy Rudoff}.} \bibinfo{year}{2017}\natexlab{}.
\newblock \showarticletitle{{NOVA-Fortis: A Fault-Tolerant Non-Volatile Main
  Memory File System}}. In \bibinfo{booktitle}{\emph{Proc. SOSP}}.
  \bibinfo{publisher}{{ACM}}, \bibinfo{pages}{478--496}.
\newblock
\urldef\tempurl%
\url{https://doi.org/10.1145/3132747.3132761}
\showDOI{\tempurl}


\bibitem[\protect\citeauthoryear{Xu, Koo, Kumar, Bai, Mitra, Misailovic, and
  Bagchi}{Xu et~al\mbox{.}}{2018b}]%
        {DBLP:conf/usenix/XuKKBMMB18}
\bibfield{author}{\bibinfo{person}{Ran Xu}, \bibinfo{person}{Jinkyu Koo},
  \bibinfo{person}{Rakesh Kumar}, \bibinfo{person}{Peter Bai},
  \bibinfo{person}{Subrata Mitra}, \bibinfo{person}{Sasa Misailovic}, {and}
  \bibinfo{person}{Saurabh Bagchi}.} \bibinfo{year}{2018}\natexlab{b}.
\newblock \showarticletitle{{VideoChef: Efficient Approximation for Streaming
  Video Processing Pipelines}}. In \bibinfo{booktitle}{\emph{Proc. ATC}}.
  \bibinfo{publisher}{{USENIX} Association}, \bibinfo{pages}{43--56}.
\newblock
\urldef\tempurl%
\url{https://www.usenix.org/conference/atc18/presentation/xu-ran}
\showURL{%
\tempurl}


\bibitem[\protect\citeauthoryear{Yan, Zhang, Wang, Strauss, and Ceze}{Yan
  et~al\mbox{.}}{2017}]%
        {DBLP:conf/hotstorage/YanZWSC17}
\bibfield{author}{\bibinfo{person}{Eddie~Q. Yan}, \bibinfo{person}{Kaiyuan
  Zhang}, \bibinfo{person}{Xi Wang}, \bibinfo{person}{Karin Strauss}, {and}
  \bibinfo{person}{Luis Ceze}.} \bibinfo{year}{2017}\natexlab{}.
\newblock \showarticletitle{{Customizing Progressive JPEG for Efficient Image
  Storage}}. In \bibinfo{booktitle}{\emph{Proc. HotStorage}}.
\newblock
\urldef\tempurl%
\url{https://www.usenix.org/conference/hotstorage17/program/presentation/yan}
\showURL{%
\tempurl}


\bibitem[\protect\citeauthoryear{Yang, Lee, Kim, Cho, Lee, and Yu}{Yang
  et~al\mbox{.}}{2007}]%
        {DBLP:conf/iscas/YangLKCLY07}
\bibfield{author}{\bibinfo{person}{Byung{-}Do Yang}, \bibinfo{person}{Jae{-}Eun
  Lee}, \bibinfo{person}{Jang{-}Su Kim}, \bibinfo{person}{Junghyun Cho},
  \bibinfo{person}{Seung{-}Yun Lee}, {and} \bibinfo{person}{Byoung{-}Gon Yu}.}
  \bibinfo{year}{2007}\natexlab{}.
\newblock \showarticletitle{{A Low Power Phase-Change Random Access Memory
  using a Data-Comparison Write Scheme}}. In \bibinfo{booktitle}{\emph{Proc.
  ISCAS}}. \bibinfo{publisher}{{IEEE}}, \bibinfo{pages}{3014--3017}.
\newblock
\urldef\tempurl%
\url{https://doi.org/10.1109/ISCAS.2007.377981}
\showDOI{\tempurl}


\bibitem[\protect\citeauthoryear{Yazdanbakhsh, Mahajan, Esmaeilzadeh, and
  Lotfi{-}Kamran}{Yazdanbakhsh et~al\mbox{.}}{2017}]%
        {DBLP:journals/dt/YazdanbakhshMEL17}
\bibfield{author}{\bibinfo{person}{Amir Yazdanbakhsh}, \bibinfo{person}{Divya
  Mahajan}, \bibinfo{person}{Hadi Esmaeilzadeh}, {and} \bibinfo{person}{Pejman
  Lotfi{-}Kamran}.} \bibinfo{year}{2017}\natexlab{}.
\newblock \showarticletitle{{AxBench: A Multiplatform Benchmark Suite for
  Approximate Computing}}.
\newblock \bibinfo{journal}{\emph{{IEEE} Design {\&} Test}}
  \bibinfo{volume}{34}, \bibinfo{number}{2} (\bibinfo{year}{2017}),
  \bibinfo{pages}{60--68}.
\newblock
\urldef\tempurl%
\url{https://doi.org/10.1109/MDAT.2016.2630270}
\showDOI{\tempurl}


\bibitem[\protect\citeauthoryear{Young, Kariyappa, and Qureshi}{Young
  et~al\mbox{.}}{2018}]%
        {DBLP:journals/corr/abs-1807-07685}
\bibfield{author}{\bibinfo{person}{Vinson Young}, \bibinfo{person}{Sanjay
  Kariyappa}, {and} \bibinfo{person}{Moinuddin~K. Qureshi}.}
  \bibinfo{year}{2018}\natexlab{}.
\newblock \showarticletitle{{CRAM: Efficient Hardware-Based Memory Compression
  for Bandwidth Enhancement}}.
\newblock \bibinfo{journal}{\emph{CoRR}}  \bibinfo{volume}{abs/1807.07685}
  (\bibinfo{year}{2018}).
\newblock
\showeprint[arxiv]{1807.07685}
\urldef\tempurl%
\url{http://arxiv.org/abs/1807.07685}
\showURL{%
\tempurl}


\bibitem[\protect\citeauthoryear{Yue and Zhu}{Yue and Zhu}{2013}]%
        {DBLP:conf/hpca/YueZ13}
\bibfield{author}{\bibinfo{person}{Jianhui Yue} {and} \bibinfo{person}{Yifeng
  Zhu}.} \bibinfo{year}{2013}\natexlab{}.
\newblock \showarticletitle{{Accelerating Write by Exploiting PCM
  Asymmetries}}. In \bibinfo{booktitle}{\emph{Proc. HPCA}}.
  \bibinfo{publisher}{{IEEE} Computer Society}, \bibinfo{pages}{282--293}.
\newblock
\urldef\tempurl%
\url{https://doi.org/10.1109/HPCA.2013.6522326}
\showDOI{\tempurl}


\bibitem[\protect\citeauthoryear{Zhao, Xue, Chi, and Zhao}{Zhao
  et~al\mbox{.}}{2017}]%
        {DBLP:conf/iccad/ZhaoXCZ17}
\bibfield{author}{\bibinfo{person}{Hengyu Zhao}, \bibinfo{person}{Linuo Xue},
  \bibinfo{person}{Ping Chi}, {and} \bibinfo{person}{Jishen Zhao}.}
  \bibinfo{year}{2017}\natexlab{}.
\newblock \showarticletitle{{Approximate Image Storage with Multi-level Cell
  STT-MRAM Main Memory}}. In \bibinfo{booktitle}{\emph{Proc. ICCAD}}.
  \bibinfo{pages}{268--275}.
\newblock
\urldef\tempurl%
\url{https://doi.org/10.1109/ICCAD.2017.8203788}
\showDOI{\tempurl}


\bibitem[\protect\citeauthoryear{Zhou, Zhao, Yang, and Zhang}{Zhou
  et~al\mbox{.}}{2009}]%
        {DBLP:conf/isca/ZhouZYZ09}
\bibfield{author}{\bibinfo{person}{Ping Zhou}, \bibinfo{person}{Bo Zhao},
  \bibinfo{person}{Jun Yang}, {and} \bibinfo{person}{Youtao Zhang}.}
  \bibinfo{year}{2009}\natexlab{}.
\newblock \showarticletitle{{A Durable and Energy Efficient Main Memory Using
  Phase Change Memory Technology}}. In \bibinfo{booktitle}{\emph{Proc. ISCA}}.
  \bibinfo{publisher}{{ACM}}, \bibinfo{pages}{14--23}.
\newblock
\urldef\tempurl%
\url{https://doi.org/10.1145/1555754.1555759}
\showDOI{\tempurl}


\bibitem[\protect\citeauthoryear{Zuo, Hua, and Wu}{Zuo et~al\mbox{.}}{2018}]%
        {DBLP:conf/osdi/Zuo0W18}
\bibfield{author}{\bibinfo{person}{Pengfei Zuo}, \bibinfo{person}{Yu Hua},
  {and} \bibinfo{person}{Jie Wu}.} \bibinfo{year}{2018}\natexlab{}.
\newblock \showarticletitle{{Write-Optimized and High-Performance Hashing Index
  Scheme for Persistent Memory}}. In \bibinfo{booktitle}{\emph{Proc. OSDI}}.
  \bibinfo{publisher}{{USENIX} Association}, \bibinfo{pages}{461--476}.
\newblock
\urldef\tempurl%
\url{https://www.usenix.org/conference/osdi18/presentation/zuo}
\showURL{%
\tempurl}


\end{thebibliography}

\end{document}